\documentstyle[12pt]{article}
\newread\epsffilein    
\newif\ifepsffileok    
\newif\ifepsfbbfound   
\newif\ifepsfverbose   
\newdimen\epsfxsize    
\newdimen\epsfysize    
\newdimen\epsftsize    
\newdimen\epsfrsize    
\newdimen\epsftmp      
\newdimen\pspoints     
\def\epsfbox#1{\global\def\epsfllx{72}\global\def\epsflly{72}%
   \global\def\epsfurx{540}\global\def\epsfury{720}%
   \def\lbracket{[}\def\testit{#1}\ifx\testit\lbracket
   \let\next=\epsfgetlitbb\else\let\next=\epsfnormal\fi\next{#1}}%
\def\epsfgetlitbb#1#2 #3 #4 #5]#6{\epsfgrab #2 #3 #4 #5 .\\%
   \epsfsetgraph{#6}}%
\def\epsfnormal#1{\epsfgetbb{#1}\epsfsetgraph{#1}}%
\def\epsfgetbb#1{%
\openin\epsffilein=#1
\ifeof\epsffilein\errmessage{I couldn't open #1, will ignore it}\else
   {\epsffileoktrue \chardef\other=12
    \def\do##1{\catcode`##1=\other}\dospecials \catcode`\ =10
    \loop
       \read\epsffilein to \epsffileline
       \ifeof\epsffilein\epsffileokfalse\else
          \expandafter\epsfaux\epsffileline:. \\%
       \fi
   \ifepsffileok\repeat
   \ifepsfbbfound\else
    \ifepsfverbose\message{No bounding box comment in #1; using defaults}\fi\fi
   }\closein\epsffilein\fi}%
\def\epsfsetgraph#1{%
   \epsfrsize=\epsfury\pspoints
   \advance\epsfrsize by-\epsflly\pspoints
   \epsftsize=\epsfurx\pspoints
   \advance\epsftsize by-\epsfllx\pspoints
   \epsfxsize\epsfsize\epsftsize\epsfrsize
   \ifnum\epsfxsize=0 \ifnum\epsfysize=0
      \epsfxsize=\epsftsize \epsfysize=\epsfrsize
     \else\epsftmp=\epsftsize \divide\epsftmp\epsfrsize
       \epsfxsize=\epsfysize \multiply\epsfxsize\epsftmp
       \multiply\epsftmp\epsfrsize \advance\epsftsize-\epsftmp
       \epsftmp=\epsfysize
       \loop \advance\epsftsize\epsftsize \divide\epsftmp 2
       \ifnum\epsftmp>0
          \ifnum\epsftsize<\epsfrsize\else
             \advance\epsftsize-\epsfrsize \advance\epsfxsize\epsftmp \fi
       \repeat
     \fi
   \else\epsftmp=\epsfrsize \divide\epsftmp\epsftsize
     \epsfysize=\epsfxsize \multiply\epsfysize\epsftmp   
     \multiply\epsftmp\epsftsize \advance\epsfrsize-\epsftmp
     \epsftmp=\epsfxsize
     \loop \advance\epsfrsize\epsfrsize \divide\epsftmp 2
     \ifnum\epsftmp>0
        \ifnum\epsfrsize<\epsftsize\else
           \advance\epsfrsize-\epsftsize \advance\epsfysize\epsftmp \fi
     \repeat     
   \fi
   \ifepsfverbose\message{#1: width=\the\epsfxsize, height=\the\epsfysize}\fi
   \epsftmp=10\epsfxsize \divide\epsftmp\pspoints
   \vbox to\epsfysize{\vfil\hbox to\epsfxsize{%
      \includegraphics{#1}%
      \hfil}}%
\epsfxsize=0pt\epsfysize=0pt}%

{\catcode`\%=12 \global\let\epsfpercent=
\long\def\epsfaux#1#2:#3\\{\ifx#1\epsfpercent
   \def\testit{#2}\ifx\testit\epsfbblit
      \epsfgrab #3 . . . \\%
      \epsffileokfalse
      \global\epsfbbfoundtrue
   \fi\else\ifx#1\par\else\epsffileokfalse\fi\fi}%
\def\epsfgrab #1 #2 #3 #4 #5\\{%
   \global\def\epsfllx{#1}\ifx\epsfllx\empty
      \epsfgrab #2 #3 #4 #5 .\\\else
   \global\def\epsflly{#2}%
   \global\def\epsfurx{#3}\global\def\epsfury{#4}\fi}%
\def\epsfsize#1#2{\epsfxsize}
\let\epsffile=\epsfbox

\def\ltap{\raisebox{-.4ex}{\rlap{$\sim$}} \raisebox{.4ex}{$<$}}

\newlength{\dinwidth}
\newlength{\dinmargin}
\setlength{\dinwidth}{21.0cm}
\textheight24.2cm \textwidth17.0cm
\setlength{\dinmargin}{\dinwidth}
\addtolength{\dinmargin}{-\textwidth}
\setlength{\dinmargin}{0.5\dinmargin}
\oddsidemargin -1.0in
\addtolength{\oddsidemargin}{\dinmargin}
 
\setlength{\evensidemargin}{\oddsidemargin}
\setlength{\marginparwidth}{0.9\dinmargin}
\marginparsep 8pt \marginparpush 5pt
\topmargin -42pt
\headheight 12pt
\headsep 30pt 
\parskip 3mm plus 2mm minus 2mm
\parindent 5mm
\renewcommand{\arraystretch}{1.3}
\renewcommand{\thefootnote}{\arabic{footnote}}

\newcommand{\bfig}{\begin{figure}}
\newcommand{\efig}{\end{figure}}
\newcommand{\bcen}{\begin{center}}
\newcommand{\ecen}{\end{center}}
\newcommand{\beq}{\begin{equation}}
\newcommand{\eeq}{\end{equation}}
\newcommand{\btabu}{\begin{tabular}}
\newcommand{\etabu}{\end{tabular}}
\newcommand{\btabl}{\begin{table}}
\newcommand{\etabl}{\end{table}}

\newcommand {\pom}  {I\hspace{-0.2em}P}
\newcommand {\alphapom} {\mbox{$\alpha_{_{\pom}}$}}

\def\lsim{\mathrel{\rlap{\lower4pt\hbox{\hskip1pt$\sim$}}
    \raise1pt\hbox{$<$}}}         
\def\gsim{\mathrel{\rlap{\lower4pt\hbox{\hskip1pt$\sim$}}
    \raise1pt\hbox{$>$}}}         

\begin{document}
\title{\bf 
Elastic and Proton-Dissociative {\boldmath $\rho^0$} 
Photoproduction at HERA}
\vspace{5cm}
\author{\Large ZEUS Collaboration} 
\date{}

\maketitle
\thispagestyle{empty}
\vspace{3cm}
\begin{abstract}
Elastic and proton-dissociative $\rho^0$ photoproduction
($\gamma p \rightarrow \rho^0 p$, $\gamma p \rightarrow \rho^0 N$, 
respectively, with $\rho^0 \rightarrow \pi^+\pi^-$)
has been studied in $ep$ interactions at HERA for 
photon-proton centre-of-mass energies in the range $50<W<100$~GeV and
for $|t|<0.5$~GeV$^2$, where $t$ is the square of the four-momentum transfer 
at the proton vertex; the results on the proton-dissociative reaction are
presented for masses of the dissociated proton system in the range
$M_N^2<0.1 W^2$. 
For the elastic process, the $\pi^+\pi^-$ 
invariant mass spectrum has been investigated as a function of $t$.
As in fixed target experiments, the $\rho^0$ resonance shape is 
asymmetric; this asymmetry decreases with increasing $|t|$,
as expected in 
models in which the asymmetry is ascribed to the interference of resonant and 
non-resonant $\pi^+\pi^-$ production. 
The cross section has been studied as a function of $W$; 
a fit to the resonant part with the form $W^a$ gives
$a=0.16\pm 0.06~\mbox{(stat.)}~^{+0.11}_{-0.15}~\mbox{(syst.)}$. 
The resonant part of the $\gamma p \rightarrow \pi^+\pi^- p$ cross section 
is $11.2\pm 0.1~\mbox{(stat.)}~^{+1.1}_{-1.2}~\mbox{(syst.)}$~$\mu$b at 
$\langle W\rangle=71.7$~GeV.
The $t$ dependence of the cross section can be
described by a function of the type  
$A_{\rho} \exp{(-b_{\rho} |t| + c_{\rho} t^2)}$ with 
$b_{\rho}= 10.9 \pm 0.3$~(stat.)~$^{+1.0}_{-0.5}$~(syst.)~GeV$^{-2}$ and 
$c_{\rho}= 2.7 \pm 0.9$~(stat.)~$^{+1.9}_{-1.7}$~(syst.)~GeV$^{-4}$.
The $t$ dependence has also been studied as a function of $W$ and a 
value of the slope of the pomeron trajectory $\alphapom^{\prime}=
0.23~\pm 0.15~\mbox{(stat.)}~^{+0.10}_{-0.07}~\mbox{(syst.)}$~GeV$^{-2}$
has been deduced.
The $\rho^0$ spin density matrix elements $r^{04}_{00}$,
$r^{04}_{1-1}$ and $\Re{e[r^{04}_{10}]}$ have been measured and found to be 
consistent with expectations based on $s$-channel helicity conservation.
For proton-dissociative $\pi^+\pi^-$ photoproduction in the $\rho^0$ 
mass range, the distributions of the 
two-pion invariant mass, $W$ and the polar and azimuthal angles of the pions 
in the helicity frame  are the same within errors as 
those for the
elastic process. 
The $t$ distribution has been fitted to an exponential function with a 
slope parameter 
$5.8 \pm~0.3~\mbox{(stat.)} \pm 0.5~\mbox{(syst.)}$~GeV$^{-2}$.
The ratio of the elastic to proton-dissociative $\rho^0$ photoproduction 
cross section is $2.0~\pm~0.2~\mbox{(stat.)}~\pm 0.7~\mbox{(syst.)}$.
\end{abstract}

\vspace{-24cm}
\begin{flushleft}
\tt DESY 97-237 \\
November 1997 \\
\end{flushleft}

\thispagestyle{empty}
\newpage

%
%
%
%
\topmargin-1.cm                                                                                    
\evensidemargin-0.3cm                                                                              
\oddsidemargin-0.3cm                                                                               
\textwidth 16.cm                                                                                   
\textheight 680pt                                                                                  
\parindent0.cm                                                                                     
\parskip0.3cm plus0.05cm minus0.05cm                                                               
\def\3{\ss}                                                                                        
\newcommand{\address}{ }                                                                           
\renewcommand{\author}{ }                                                                          
\pagenumbering{Roman}                                                                              
                                                   %
\begin{center}                                                                                     
{                      \Large  The ZEUS Collaboration              }                               
\end{center}                                                                                       
  J.~Breitweg,                                                                                     
  M.~Derrick,                                                                                      
  D.~Krakauer,                                                                                     
  S.~Magill,                                                                                       
  D.~Mikunas,                                                                                      
  B.~Musgrave,                                                                                     
  J.~Repond,                                                                                       
  R.~Stanek,                                                                                       
  R.L.~Talaga,                                                                                     
  R.~Yoshida,                                                                                      
  H.~Zhang  \\                                                                                     
 {\it Argonne National Laboratory, Argonne, IL, USA}~$^{p}$
% ...
\newpage                                                                                           
$^{\    1}$ also at IROE Florence, Italy \\                                                        
$^{\    2}$ now at Univ. of Salerno and INFN Napoli, Italy \\                                      
% ...
$^{\    4}$ now at E.S.R.F., BP220, F-38043 Grenoble, France \\                                    
$^{\    5}$ supported by Worldlab, Lausanne, Switzerland \\                                        
% ...
$^{\    7}$ retired \\                                                                             
% ...
$^{\    9}$ now at Dongshin University, Naju, Korea \\                                             
% ...
$^{  25}$ now at ATLAS Collaboration, Univ. of Munich \\                                           
% ...
$^{  34}$ partially supported by the Foundation for German-Russian Collaboration                   
DFG-RFBR \\ \hspace*{3.5mm} (grant no. 436 RUS 113/248/3 and no. 436 RUS 113/248/2)\\              
% ...
                                                           %
                                                           %
\newpage   
                                                           %
                                                           %
\begin{tabular}[h]{rp{14cm}}                                                                       
$^{a}$ &  supported by the Natural Sciences and Engineering Research                               
          Council of Canada (NSERC)  \\                                                            
$^{b}$ &  supported by the FCAR of Qu\'ebec, Canada  \\                                            
$^{c}$ &  supported by the German Federal Ministry for Education and                               
          Science, Research and Technology (BMBF), under contract                                  
          numbers 057BN19P, 057FR19P, 057HH19P, 057HH29P, 057SI75I \\                              
$^{d}$ &  supported by the MINERVA Gesellschaft f\"ur Forschung GmbH,                              
          the German Israeli Foundation, and the U.S.-Israel Binational                            
          Science Foundation \\                                                                    
$^{e}$ &  supported by the German Israeli Foundation, and                                          
          by the Israel Science Foundation                                                         
  \\                                                                                               
$^{f}$ &  supported by the Italian National Institute for Nuclear Physics                          
          (INFN) \\                                                                                
$^{g}$ &  supported by the Japanese Ministry of Education, Science and                             
          Culture (the Monbusho) and its grants for Scientific Research \\                         
$^{h}$ &  supported by the Korean Ministry of Education and Korea Science                          
          and Engineering Foundation  \\                                                           
$^{i}$ &  supported by the Netherlands Foundation for Research on                                  
          Matter (FOM) \\                                                                          
$^{j}$ &  supported by the Polish State Committee for Scientific                                   
          Research, grant No.~115/E-343/SPUB/P03/002/97, 2P03B10512,                               
          2P03B10612, 2P03B14212, 2P03B10412 \\                                                    
$^{k}$ &  supported by the Polish State Committee for Scientific                                   
          Research (grant No. 2P03B08308) and Foundation for                                       
          Polish-German Collaboration  \\                                                          
$^{l}$ &  partially supported by the German Federal Ministry for                                   
          Education and Science, Research and Technology (BMBF)  \\                                
$^{m}$ &  supported by the Fund for Fundamental Research of Russian Ministry                       
          for Science and Edu\-cation and by the German Federal Ministry for                       
          Education and Science, Research and Technology (BMBF) \\                                 
$^{n}$ &  supported by the Spanish Ministry of Education                                           
          and Science through funds provided by CICYT \\                                           
$^{o}$ &  supported by the Particle Physics and                                                    
          Astronomy Research Council \\                                                            
$^{p}$ &  supported by the US Department of Energy \\                                              
$^{q}$ &  supported by the US National Science Foundation \\                                       
\end{tabular}                                                                                      

\newpage

\setlength{\dinwidth}{21.0cm}
\textheight24.2cm \textwidth17.0cm
\setlength{\dinmargin}{\dinwidth}
\addtolength{\dinmargin}{-\textwidth}
\setlength{\dinmargin}{0.5\dinmargin}
\oddsidemargin -1.0in
\addtolength{\oddsidemargin}{\dinmargin}
 
\setlength{\evensidemargin}{\oddsidemargin}
\setlength{\marginparwidth}{0.9\dinmargin}
\marginparsep 8pt \marginparpush 5pt
\topmargin -42pt
\headheight 12pt
\headsep 30pt 
\parskip 3mm plus 2mm minus 2mm
\parindent 5mm
\renewcommand{\arraystretch}{1.3}
\renewcommand{\thefootnote}{\arabic{footnote}}
\pagenumbering{arabic} 
\setcounter{page}{1}

\section{Introduction}

Elastic photoproduction of $\rho^0$ mesons, $\gamma p \rightarrow \rho^0 p$,
has been studied in fixed target 
experiments at photon-proton centre-of-mass energies $W$ 
up to 20 GeV~\cite{bauer}-\cite{omega} and at 
HERA~\cite{rho93}-\cite{lps_rho} for $W$ up to approximately 200 GeV. 
In both cases the reaction exhibits the features of a soft 
diffractive process, namely a weak energy dependence and 
a differential cross section 
$d\sigma/dt \propto \exp{(bt)}$ at low $|t|$ values, where 
$t$ is the squared four-momentum exchanged between the photon and the proton. 
These features, typical of elastic hadron-hadron interactions,
are consistent with the expectations of the Vector Meson Dominance 
model (VDM)~\cite{sakurai} in which the photon is assumed to 
fluctuate into a vector meson before scattering from the proton.
In elastic $\rho^0$ photoproduction the photon thus appears to behave 
like an ordinary hadron which interacts elastically with the proton. 
Many aspects of $\rho^0$ photoproduction 
remain however to be clarified~--~among them, the $W$ dependence 
of the cross section, the origin of the asymmetric $\rho^0$ resonance shape 
and the extent to which the helicity of the photon is transferred to 
the vector meson. Perturbative QCD 
calculations which have been able to succesfully describe the photoproduction 
of $J/\psi$ mesons~\cite{jpsi} are not strictly applicable to $\rho^0$ 
photoproduction at low $|t|$. 
In general, photoproduction of 
$\rho^0$ mesons at HERA may offer a means of investigating the nature of 
soft hadronic interactions as well as the hadronic features of the photon. 

Little is known about proton-dissociative $\rho^0$ production
with real photons, $\gamma p \to \rho^0 N$, where $N$ is a state of mass 
$M_N$ into which the proton diffractively dissociates. 
Data exist for the virtual photon case: the H1 Collaboration at HERA has
recently investigated proton-dissociative $\rho^0$ production for 
photon virtualities $Q^2 >7$~GeV$^2$~\cite{h1_pdiss}.
The H1 results indicate that the cross section for this process has the same
dependence on $Q^2$ and $W$ and the same helicity structure as the 
elastic reaction. 
These observations support the hypothesis of 
factorisation of the diffractive vertex \cite{reviews}, which
has been extensively studied in hadron-hadron reactions
(see e.g.~\cite{reviews}-\cite{CDF}).
The H1 data also show that the $t$ distribution is exponential but shallower 
than that for the elastic case. 
In photoproduction, proton-dissociative production of $\rho^0$ 
mesons can provide yet another way to 
study diffraction, the hadronic properties of the photon and the nature of soft 
hadronic processes. In conjunction with the elastic reaction, it can provide 
a test of factorisation.
Moreover, a detailed understanding of proton-diffractive dissociation
is mandatory for the study of the elastic reaction, for which it is the
main source of background when the scattered proton is not measured.

\bigskip
This paper describes a measurement of $\rho^0$ 
photoproduction in the elastic 
and proton-dissocia\-tive 
reactions. 
The measurement was performed using data collected in 1994 by the ZEUS 
experiment at HERA for the processes $ep \rightarrow e \pi^+ \pi^- p$ and
$ep \rightarrow e \pi^+ \pi^-  N$ at small photon virtualities, 
$Q^2 \lsim 4$~GeV$^2$. The symbol $e$ indicates positrons.

For the data presented here the scattered positron
was not detected. The scattered proton was measured only for a subsample of
the data. In general the relevant kinematic quantities were determined 
from the measured three-momenta of the two pions from the $\rho^0$ decay.

With respect to the ZEUS 1993 data~\cite{rho93}, the present results 
on elastic $\rho^0$ photoproduction feature larger statistics,
a wider $W$ range and  smaller systematic uncertainties.
Two of the main contributions to the uncertainties of the 1993 
results were significantly reduced: the calorimeter trigger 
efficiency was evaluated directly from the data and the contamination of 
proton-dissociative events was determined by using a subsample 
of the data in which elastic events 
were unambiguously selected by detecting the final state proton.

The larger statistics and wider kinematic range allowed 
the study of the $\pi^+ \pi^-$ mass spectrum as a function of 
$t$, $W$ and the decay pions' polar and azimuthal angles
in the helicity frame. The shape of the $\pi^+ \pi^-$ mass spectrum in the reaction 
$ep \rightarrow e \pi^+ \pi^- p$ is interesting since, in the framework of 
the S\"oding model~\cite{soeding}, it depends on 
the interference between resonant $\rho^0$ and non-resonant 
$\pi^+ \pi^-$ production.

The cross section for elastic $\rho^0$ 
photoproduction, $\sigma_{\gamma p \to \rho^0 p}$, was extracted 
as a function of 
$W$. The $W$ dependence of the cross section, in Regge theory~\cite{regge}, 
is related to 
the intercept $\alphapom(0)$ of the pomeron trajectory exchanged 
between the proton and the hadronic fluctuation of the photon. 

The differential cross section $d\sigma/d|t|$ was determined and 
its shape studied as a function of $W$. Regge theory predicts
that the slope of the exponential $t$ distribution becomes
increasingly steep with increasing $W$; the rate of change of 
the $t$ slope with $W$, at high values of $W$, is related to the slope of the 
pomeron trajectory, $\alphapom^{\prime}$.

The decay pion angular distributions in the helicity frame were 
studied and the $\rho^0$ spin density matrix elements $r^{04}_{00}$,
$r^{04}_{1-1}$ and $\Re{e[r^{04}_{10}]}$ determined. 
The behaviour of these 
matrix elements as a function of the two-pion invariant mass 
$M_{\pi \pi}$ and of $W$ was investigated. This made it possible to 
test the validity of $s$-channel helicity conservation 
(SCHC). 

\bigskip
We also present results on $\rho^0$ photoproduction with diffractive 
dissociation of the proton in the range
$M_N^2<0.1 W^2$. The limit was chosen 
following refs.~\cite{reviews,chapin} and corresponds to the region where
diffractive interactions dominate. 
The distributions of $M_{\pi \pi}$, $W$, the polar and azimuthal angles
of the pions in the helicity frame and the
$t$ dependence of proton-dissociative $\rho^0$ 
photoproduction were studied. 
The ratio of the cross sections for 
elastic and proton-dissociative $\rho^0$ photoproduction was determined.

\bigskip
Finally, we used the data on the reaction 
$ep \rightarrow e \pi^+ \pi^- p$ to evaluate the pion-proton 
total cross section in a model dependent way. As mentioned earlier, the
shape of the $\pi^+ \pi^-$ mass spectrum is sensitive to the interference 
between resonant $\rho^0$ and non-resonant $\pi^+ \pi^-$ 
production. In the latter case, one (or both) of the pions interacts with 
the proton. The mass spectrum thus depends on the pion-proton 
cross section. In the framework of a calculation by Ryskin and 
Shabelski~\cite{misha_yuly}, we determined this cross section at a 
pion-proton centre-of-mass energy of the order of 50 GeV, beyond the 
reach of any existing pion beam.

\section{Experimental set-up}
\label{setup}

\subsection{HERA}

The data discussed here were collected in 1994 using the 
HERA collider which operated with 820~GeV protons and 27.5~GeV positrons. The 
proton and positron beams each contained 153 colliding bunches, 
together with 17 additional unpaired proton and 15 unpaired positron 
bunches. These 
additional bunches were used for background studies. 

\subsection{The ZEUS detector}

A detailed description of the ZEUS detector can be found 
elsewhere~\cite{detector_a,detector_b}. 
The components which are most relevant for this analysis are briefly 
discussed below.

Charged particles are
tracked by the inner tracking detectors which operate in a
magnetic field of 1.43 T provided by a thin superconducting solenoid.
Immediately surrounding the beam pipe is the vertex detector (VXD), 
a drift chamber with 120 radial cells, each with 12 sense wires~\cite{vxd}. 
It is surrounded by the central tracking detector (CTD), which consists 
of 72 cylindrical drift chamber layers, organised into 9 
superlayers covering the polar angle region\footnote{The 
coordinate system used in this paper has the 
$Z$ axis pointing in
the proton beam direction, hereafter referred to as ``forward'',
the $X$ axis pointing horizontally towards the centre of HERA and
the $Y$ axis pointing upwards. The polar angle
$\theta$ is defined with respect to the  $Z$ direction.} 
$15^\circ < \theta < 164^\circ$~\cite{ctd}. The transverse momentum 
resolution for tracks traversing all superlayers is 
$\sigma(p_T)/p_T \simeq \sqrt{(0.005p_T)^2+(0.016)^2}$, with $p_T$ in GeV.
The Rear Tracking Detector (RTD) consists of a planar drift 
chamber with three layers of drift cells with the wires oriented 
at $0^\circ$, $+60^\circ$ and $-60^\circ$ with respect to the horizontal plane;
polar angles between $160^\circ$ and $170^\circ$ are covered~\cite{rtd}.

The high resolution uranium-scintillator calorimeter (CAL) \cite{CAL}  
consists of three parts: the forward (FCAL), the barrel (BCAL) and the rear
calorimeter (RCAL); they cover the polar angle regions $2.6^\circ$ to 
$36.7^\circ$, $36.7^\circ$ to $129.1^\circ$, and $129.1^\circ$
to $176.2^\circ$, respectively. 
Each part is subdivided transversely into towers.
The towers are segmented 
longitudinally into one electromagnetic section (EMC) and one (in RCAL)
or two (in BCAL and FCAL) hadronic sections (HAC). These sections are further
subdivided into cells; each cell is viewed by two photomultiplier tubes.
The CAL energy resolution, as measured under test beam conditions,
is $\sigma_E/E=0.18/\sqrt{E}$ for electrons and $\sigma_E/E=0.35/\sqrt{E}$ 
for hadrons ($E$ in GeV). 

The Veto Wall, the C5 counter and the small angle rear tracking detector 
(SRTD)~\cite{srtd}
all consist of scintillation counters and
are located at $Z =- 730$~cm, $Z =- 315$~cm and $Z =- 150$~cm, 
respectively. 
Particles which are generated by interactions of protons with 
residual gas molecules in the beam pipe (proton ``beam-gas" events) 
upstream of the nominal $ep$ interaction point reach the RCAL, the 
Veto Wall, the SRTD and C5 at different times 
than particles originating from the nominal $ep$ interaction point.
Proton beam-gas events are thus rejected by timing measurements
in these detectors.

The proton remnant tagger 
(PRT1)~\cite{mx} is used to tag events in which the proton diffractively 
dissociates. It consists of two layers of scintillation counters 
perpendicular to the beam and is positioned at $Z=515$~cm.
The two layers are separated by a 1~mm thick lead absorber.
Each layer is split vertically into two halves and each half
is read out by a photomultiplier tube. 
The counters have an active area of $30\,{\rm cm}\times 26\,{\rm cm}$ 
with a hole of $6.0\,{\rm cm}\times 4.5\,{\rm cm}$ 
at the centre to accommodate the HERA beam pipe.  
The PRT1 covers the range in pseudorapidity 
($\eta= - \ln{\tan{(\theta/2)}}$) from 4.3 to 5.8.

The Leading Proton Spectrometer (LPS)~\cite{lps_rho} detects
charged particles scattered at small angles and carrying a substantial 
fraction, $x_L$, of the incoming proton momentum; these particles remain 
in the beam pipe and their trajectory is measured by a system of position 
sensitive silicon micro-strip detectors very close to the proton beam. The 
detectors are located in six stations, 
S1 to S6, placed along the beam line in the direction of the outgoing 
protons, at $Z=23.8$~m, 40.3~m, 44.5~m, 
63.0~m, 81.2~m and 90.0~m from the interaction point, respectively.
The track deflections induced by the magnets in the 
proton beam line are used for the momentum analysis of the scattered proton.
For the present measurement, only the stations S4, S5 and S6 were used.
With this configuration, for $x_L$ close to unity, a resolution of 0.4\% 
on the longitudinal 
momentum and 5~MeV on the transverse momentum has been achieved. 
The 
transverse momentum
resolution is however dominated by the proton beam intrinsic transverse 
momentum spread at the interaction point of $\approx 40$~MeV in the horizontal 
plane and $\approx 90$~MeV in the vertical plane.

The luminosity was determined from the rate of the Bethe-Heitler 
process, $ep \rightarrow e \gamma p$, where the photon is measured with a 
calorimeter (LUMI) located at $Z= - 107$~m in the HERA tunnel downstream of the 
interaction point in the direction of the outgoing positrons~\cite{lumi}.

\section{Kinematics}
\label{Sect:Kinem}

\begin{figure}
\vspace{-2.4cm}
\begin{center}
\leavevmode
\hbox{%
\epsfxsize = 15cm
\epsffile{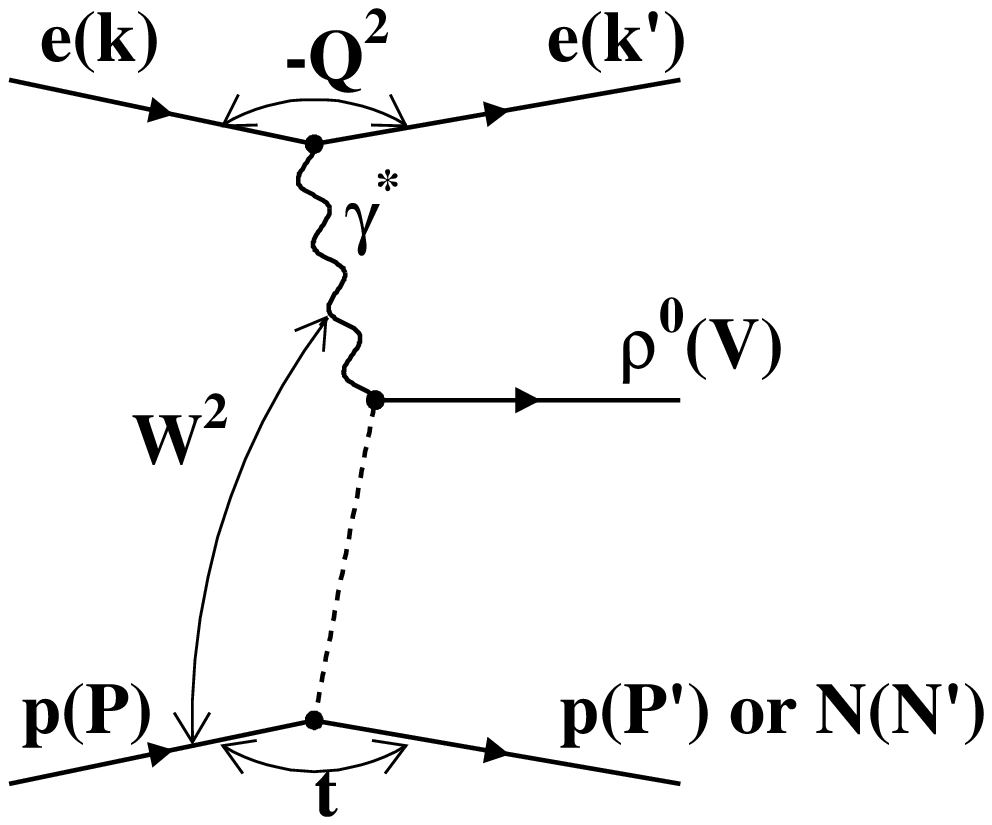}}
\end{center}
\vspace{1cm}
\caption{Elastic or proton-dissociative $\rho^0$ production in $ep$ collisions.
}
\label{reactions}
\end{figure}

The reactions under study (cf.~Fig.~\ref{reactions}) are
\begin{equation}
    e(k)p(P) \rightarrow e(k')\rho^0(V)p(P')~\mbox{and}~
    e(k)p(P) \rightarrow e(k')\rho^0(V)N(N'),
\end{equation}
where the symbols in parentheses denote the four-momenta of the 
corresponding particles (or particle system, in the case of $N$). 

The kinematics of the inclusive scattering of 
unpolarised positrons and protons is
described by the positron-proton centre-of-mass energy squared, $s$,
and any two of the following variables:
\begin{itemize}
    \item $Q^2=-q^2=-(k-k')^2$, the negative square
    of the exchanged photon's four-momentum.
    \item $y=(q\cdot P)/(k\cdot P)$, the fraction of the positron energy 
         transferred to the hadronic final state in the rest frame of the 
         initial state proton.
    \item $W^2 = (q+P)^2= -Q^2+2y(k\cdot P)+M^2_p \simeq ys$, the
         centre-of-mass energy squared of the photon-proton system,
         where $M_p$ is the proton mass.
\end{itemize}

For the exclusive reaction $e p \rightarrow e \rho^0 p$
($\rho^0\rightarrow \pi^{+}\pi^{-}$) and the proton-dissociative process
$e p \rightarrow e \rho^0 N$, the following additional variables 
are used: 
\begin{itemize}
    \item $t =(q-V)^2 = (P-P')^2$, the four-momentum transfer squared at the
    photon-$\rho^0$ vertex; for the proton-dissociative
    reaction, $t =(q-V)^2 = (P-N')^2$.
    \item The angle between the $\rho^0$ production plane (which contains
    the virtual photon and the $\rho^0$) and the positron 
         scattering plane.
    \item The polar and azimuthal angles, $\theta_h$ and $\varphi_h$, of
         the decay $\pi^+$ in the $\rho^0$ 
         helicity frame, where
the $\rho^0$ is at rest and the polar angle $\theta_h$ 
is defined as the angle between the direction opposite to that of 
the outgoing proton and 
the direction of the $\pi^+$. The azimuthal angle $\varphi_h$ is the angle between 
the decay plane and the $\rho^0$ production plane. 
     \item $x_L$, the fraction of the incoming beam momentum carried by
     the outgoing proton.
     \item For the proton-dissociative reaction, the mass $M_N$
     of the diffractively produced state $N$ is relevant. 
     In the present analysis however it was not possible to measure this 
     quantity directly and the $M_N$ range covered was obtained from 
     Monte Carlo simulations (see section~\ref{selection_inelastic}).

\end{itemize}

In the present analysis, events were selected in which the final state positron
was scattered at an angle too small to be detected in the uranium
calorimeter. 
Thus the angle between the $\rho^0$ production plane and the positron scattering
plane was not measured.
In such untagged photoproduction events, the $Q^2$ value ranges from the 
kinematic minimum $Q^2_{min} = M^2_e y^2/(1-y) \sim 10^{-9}~\rm{GeV^2}$, 
where $M_e$ is the electron mass, to the value at which the scattered positron
is observed in the uranium calorimeter,
$Q^2_{max} \approx 4 \ \rm{GeV^2}$, with a median $Q^2$ of
approximately $4 \times 10^{-6} {\rm GeV^2}$. 
Since the typical $Q^2$ is small, the photon-proton centre-of-mass energy can
be approximated by
\begin{equation}
    W^2 = 4 E_p E_e y \simeq 2 (E_{\rho} - p_{Z \rho}) E_p,
\label{Eq:W2Def}
\end{equation}
where $E_p$, $E_e$ and $E_{\rho}$ are the energies 
of the incoming proton, of the incoming positron and of the $\pi^+\pi^-$ system,
respectively;
the longitudinal momentum of the $\pi^+\pi^-$ system 
is denoted by $p_{Z \rho}$. 
Furthermore for $Q^2 = Q^2_{min}$, $t$ is given by
\begin{equation}
    {\it t} = (q-V)^2 \simeq -p^2_{T \rho},
\end{equation}
where $p_{T\rho}$ is the momentum of the $\pi^+\pi^-$ system~transverse to 
the beam axis.
Non-zero values of $Q^2$ cause $t$ to differ from $-p^2_{T\rho}$ by less 
than $Q^2$. A multiplicative correction factor determined with the 
Monte Carlo generators discussed in section~\ref{sec_montecarlo} was 
applied to the $p^2_{T\rho}$ distribution to account for this effect;
the correction was obtained by taking the ratio between the
$t$ and $p_{T \rho}^2$ distributions at the generator level (cf. e.g.~\cite{rho93}).
The correction varies from 1.13 at $p_{T \rho}^2 = 0$ to 0.62 at 
$p_{T \rho}^2 = 0.5$~GeV$^2$. The result thus obtained
is consistent with that found by using LPS tagged events~\cite{lps_rho}, 
for which $t$ is measured directly.

\section{Trigger}
\label{trigger}

ZEUS uses a three-level trigger system~\cite{detector_a,detector_b}.
For the present data, the trigger 
selected events from photoproduction of a vector meson decaying 
into two charged particles without requiring that the scattered positron be 
detected.

The first-level trigger required an energy deposit of at least 464~MeV 
in the electromagnetic section of RCAL (excluding the towers immediately 
around the beam pipe) and at least one track candidate in the CTD. 
Events with an energy deposit larger than 
1250~MeV in the FCAL towers surrounding 
the beam pipe were rejected in order to 
suppress proton beam-gas events along with a large fraction of 
photoproduction events. This cut also removes large-$M_N$ proton-dissociative
events.

At the second-level trigger, the background was reduced by using the 
measured time of the energy deposits and the summed energies from the 
calorimeter.

The full event information was available at the third-level trigger and
a simplified reconstruction procedure was used.
Tighter calorimeter timing cuts as well as algorithms to remove 
cosmic muons were applied. Exactly one  reconstructed vertex 
was demanded, with a $Z$ coordinate within $\pm 66$~cm of the nominal 
interaction point.  Furthermore, the events were required to satisfy at 
least one of the following conditions:
\begin{enumerate}
\item less than four reconstructed tracks and at least one pair with 
invariant mass less than 1.5~GeV (assuming they are pions); 
\item less than six reconstructed tracks with a total invariant 
mass less than 2.5~GeV (again assuming pions).
\end{enumerate}

\noindent
Both sets of third-level triggers were prescaled by a factor six.
An integrated luminosity of $ 2.17 \pm 0.03$~pb$^{-1}$ thus
yielded approximately 725,000 events.

No requirements were imposed on the LPS or PRT1 at the trigger level.

\begin{figure}
\vspace{-2.4cm}
\begin{center}
\leavevmode
\hbox{%
\epsfxsize = 15cm
\epsffile{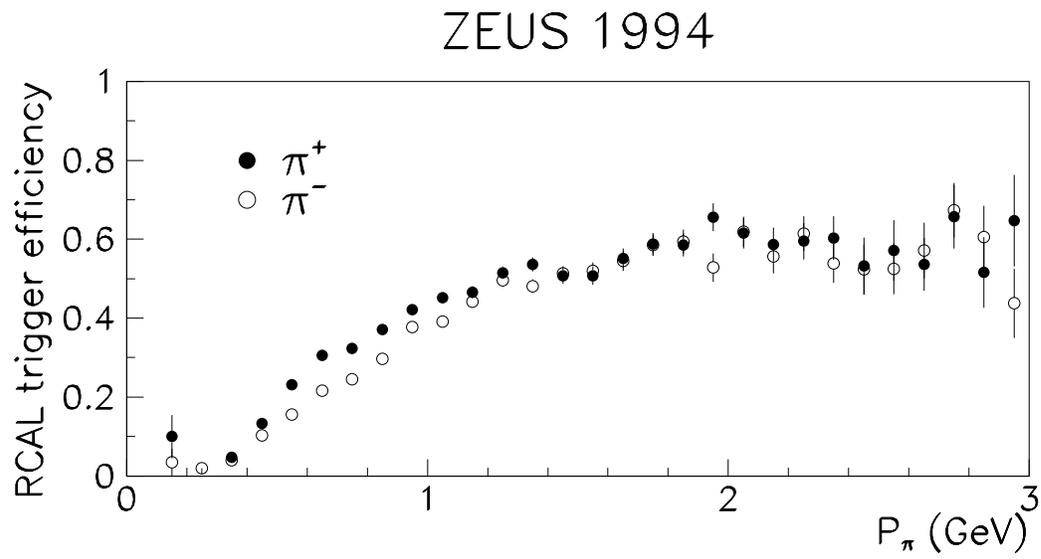}}
\end{center}
\vspace{1cm}
\caption{RCAL trigger efficiency as a function of the pion momentum $P_{\pi}$. 
The full symbols refer to positive pions and the open ones to negative pions. 
Only statistical errors are shown. 
}
\label{rcal_eff}
\end{figure}

For the present analysis, unlike what was done in~\cite{rho93,lps_rho},
the RCAL trigger efficiency at the first level was 
determined~\cite{thesis_dirk,thesis_joern} using the data rather than a 
Monte Carlo simulation. A sample of 
two-track events ($\rho^0$ candidates) was used. 
Since one of the two pions is sufficient to trigger the event,
the efficiency for RCAL to trigger on a charged pion was evaluated 
as the fraction of events in which
the second pion could have satisfied the trigger and in which it 
actually did. This was feasible
since, for a subsample of the events, it was possible to uniquely determine
which of the two pions satisfied the RCAL trigger.
The results for the efficiency were parametrised as 
a function of the momentum and polar angle of the pion, 
separately for positive and negative pions. The efficiency
was then applied as a multiplicative weight 
to each event.
Figure~\ref{rcal_eff}
shows the efficiency as a function of the momentum. The 
uncertainty is dominated by statistics. 
The same parametrisation was also used in~\cite{zeus_omega}.

\section{Event selection}

\subsection{Event selection for elastic $\rho^0$ photoproduction}
\label{selection_elastic}

The following offline requirements were imposed to select 
candidates for the reaction $ep \rightarrow e \pi^+\pi^- p$:

\begin{itemize}

\item Exactly two tracks in the CTD from particles of opposite charge, both 
associated with the reconstructed vertex.

\item The coordinates of the reconstructed vertex in the range 
$-0.5< X< 0.8$~cm, $-0.8< Y< 0.5$~cm and $-29< Z< 38$~cm
(approximately corresponding to three standard deviations of the 
vertex distribution).

\item Transverse momentum greater than 150 MeV and $|\eta|~<~2.1$
for each of the two tracks, thus restricting the data to a region of 
well understood track reconstruction efficiency.

\item Each CAL cell which is more than 40~cm (in the EMC) or 55~cm (in 
the HAC) away from the extrapolated impact position of either 
track should not have an energy deposit above a given value.
The maximum 
allowed energy deposits varied  from 160 to 240~MeV depending on the 
calorimeter part and section. 
This cut rejects events with additional particles, including 
events with the scattered positron in RCAL.

\end{itemize}
\noindent 

After applying these requirements, the pion mass was assigned to each track 
and the analysis was restricted to events reconstructed in the kinematic 
region defined by:
\begin{eqnarray}
                0.55 < & M_{\pi\pi}  & <  1.2 ~\mbox{GeV}, \nonumber \\
                       & p_{T \rho}^2       & <  0.5~\mbox{GeV}^2, \label{kin}  \\
               50    < & W & < 100   ~\mbox{GeV}. \nonumber 
\end{eqnarray}
\noindent
The restricted range in the two-pion invariant mass $M_{\pi\pi}$ 
reduces the contamination from reactions involving other mesons, in particular 
from $\phi$ production with subsequent $\phi \rightarrow K^+K^-$ decay and 
$\omega \rightarrow \pi^+\pi^-\pi^0$ production.  
The requirement on $p_{T\rho}^2$ limits the background 
from proton-dissociative $\rho^0$ production and the selected $W$ range 
restricts the data to a region of well understood acceptance.
The final sample contains 79,010 events. 

\subsection{Event selection for proton-dissociative $\rho^0$ photoproduction}
\label{selection_inelastic}

To select candidates for the reaction $ep \rightarrow e \pi^+ \pi^- N$
all the criteria discussed for the elastic events were 
applied, except for the cut on the maximum energy deposit in 
FCAL outside a region around the track impact point.
In addition, one of the following three requirements was imposed:

\begin{itemize}

\item A signal from the PRT1, tagging particles
which originate from proton dissociation.
A signal from the PRT1 was defined as a coincidence of signals consistent with 
that of at least a minimum ionising particle from 
both scintillator counter layers. 
In addition the energy deposit in the FCAL towers around the beam pipe 
was required to be less than 1.2~GeV; this was dictated by the trigger 
condition discussed above.

This sample contains 2130 events, corresponding to a luminosity of 
approximately 0.7~pb$^{-1}$ for which the PRT1 was operational.

\item An energy deposit in the FCAL towers around the beam pipe 
between 0.4~GeV and 1.2~GeV. The lower limit reduces the 
contribution from calorimeter noise; the upper one was again
a consequence of the trigger condition.
Here also the particles from the proton dissociation are tagged.

A total of 945 events was selected.

\item A proton measured in the LPS carrying a fraction of the incoming 
beam momentum $x_L<0.98$. As discussed in~\cite{lps_rho}, the $x_L$ 
spectrum measured by the LPS is characterised by a narrow peak at $x_L \approx 1$ from 
elastic events and a broad distribution for $x_L \lsim 0.98$ 
ascribed to proton-dissociative events. The cut $x_L<0.98$ 
thus tags the events in which the baryon 
from the proton dissociation is a proton and rejects elastic events, for
which $x_L$ differs from unity by $(Q^2+M_{\rho}^2 +|t|)/W^2$~\cite{lps_rho}, 
i.e. at most 0.2\% for photoproduction.

This sample contains 576 events, corresponding to a luminosity of approximately
0.9~pb$^{-1}$ for which the LPS was operational~\cite{lps_rho}.
\end{itemize}

In all cases the $M_N$ region covered is approximately $M_N \lsim 10$~GeV. This 
limit is set 
by the requirement at the first-level trigger that less than 1250~MeV be 
deposited in the FCAL towers around the beam pipe 
(cf. section~\ref{trigger}); since $M_N$ could not be measured directly,
this limit was determined by Monte Carlo studies. 

\section{Monte Carlo generators and acceptance determination}
\label{sec_montecarlo}

The reaction $ep \rightarrow e\rho^0 p$ was modelled using the 
EPSOFT~\cite{michal} generator, developed in the framework of 
HERWIG~\cite{herwig}. The generated $M_{\pi\pi}$, $W$ and $t$ 
distributions were reweighted so as to 
reproduce the measured distributions after reconstruction. Similarly
reweighted were 
the polar and azimuthal angular distributions of the decay pions in the 
helicity frame.
The effective $W$ dependence of the $\gamma p$ cross 
section was taken as $\sigma \propto W^{0.2}$. The $t$ distribution was 
generated as $A\exp{(-b|t|+ct^2)}$ with $b=11$~GeV$^{-2}$ and $c=4$~GeV$^{-4}$.  
The DIPSI~\cite{dipsi} generator was used as a cross check of the 
results obtained with EPSOFT. The LPS acceptance was determined using 
the average of DIPSI and EPSOFT.

For the simulation of the reaction $ep \rightarrow e  \rho^0 N$, 
the EPSOFT Monte Carlo was used (in the case of the PRT1 and FCAL tagged
events); for this process, the program is based 
on the assumption that the cross section for the reaction 
$\gamma p \rightarrow \rho^0 N$ is of the form:
\begin{eqnarray}
\frac{d^2\sigma}{dtdM_N^2}=\frac{1}{2} 
\frac{d\sigma_{\gamma p \rightarrow \rho^0p}}{dt}
\left(\frac{d\sigma_{pp \rightarrow pN}}{dtdM_N^2}/\frac{d\sigma_{pp \rightarrow pp}}{dt}\right),
\label{diss_1}  
\end{eqnarray}
\noindent
where the ratio 
$\frac{d\sigma_{pp\rightarrow pN}}{dtdM_N^2}/\frac{d\sigma_{pp \rightarrow pp}}{dt}$
is obtained from fits to $pp$ data~\cite{michal}.

As a cross check of the results obtained with EPSOFT, the PYTHIA 
generator~\cite{pythia} was also used (except for the proton-dissociative LPS 
tagged events for which the acceptance was determined with PYTHIA while EPSOFT 
was used as a cross check). A cross section of the form 
$d^2\sigma /dt dM_N^2 \propto e^{-b|t|}F_{sd}(M_N)/M_N^2$ is assumed in PYTHIA 
with $b=b_0+2\alphapom^{\prime} \ln{(W^2/M_N^2)}$, $b_0=2.8$~GeV$^{-2}$ and 
$\alphapom^{\prime}=0.25$~GeV$^{-2}$,
corresponding to an effective $b\simeq5$ GeV$^{-2}$ in the kinematic region
for which we present our results. The function $F_{sd}(M_N)$ enhances 
the cross section in the 
low mass resonance region and suppresses the production 
of very large masses~\cite{pythia}. A fit to the generated $M_N$ spectrum
for $10<M_N^2<200$~GeV$^2$ with a function of the type $1/M_N^n$ gives $n=2.2$.
The effect of the functions $F_{sd}(M_N)$ and $b=b(M_N)$ 
on the spectrum thus is
consistent with the result $n = 2.24 \pm 0.03$
obtained for the diffractive dissociation of the
proton in $\bar{p}p$ collisions~\cite{CDF}. 

For both EPSOFT and PYTHIA, the value of $M_N$ ranged 
between $M_p+2M_{\pi}$ and a maximum fixed by the condition discussed above,
$M_N^2/W^2 \le 0.1$~\cite{reviews,chapin}. Although the data extend down to
$M_N=M_p+M_{\pi}$, the lack of Monte Carlo events below $M_p+2M_{\pi}$
is not expected to give a significant effect~\cite{reviews}.
The shapes of the two-pion invariant 
mass distribution and the $\rho^0$ decay angular distributions 
were assumed to be the same as those of the elastic events; this assumption is
supported by the data, as discussed in section~\ref{p_diss_massetc}.

The radiation of real photons from the incoming or
outgoing positron was not simulated, nor were vacuum polarisation loops 
in the virtual photon; their effects on the cross section were 
estimated to be smaller than 2\%~\cite{kurek}.

The generated events were passed through a detailed simulation of 
the ZEUS detector and trigger. They were then subjected to 
the same reconstruction and analysis programs as the data. 
It was checked that all measured distributions were described well by 
the simulated events. The acceptance in a given bin 
was then determined as the ratio of the number of accepted
Monte Carlo events to the number generated in the selected kinematic 
range. The acceptance, calculated in this manner, accounts for the 
geometric acceptance, the detector and reconstruction efficiencies,
the detector resolution and the trigger efficiency. 
As explained in section~\ref{trigger} however, the efficiency of the RCAL 
trigger was evaluated from the data and then applied as a multiplicative weight 
to each event.

\begin{figure}
\vspace{-2.4cm}
\begin{center}
\leavevmode
\hbox{%
\epsfxsize = 15cm
\epsffile{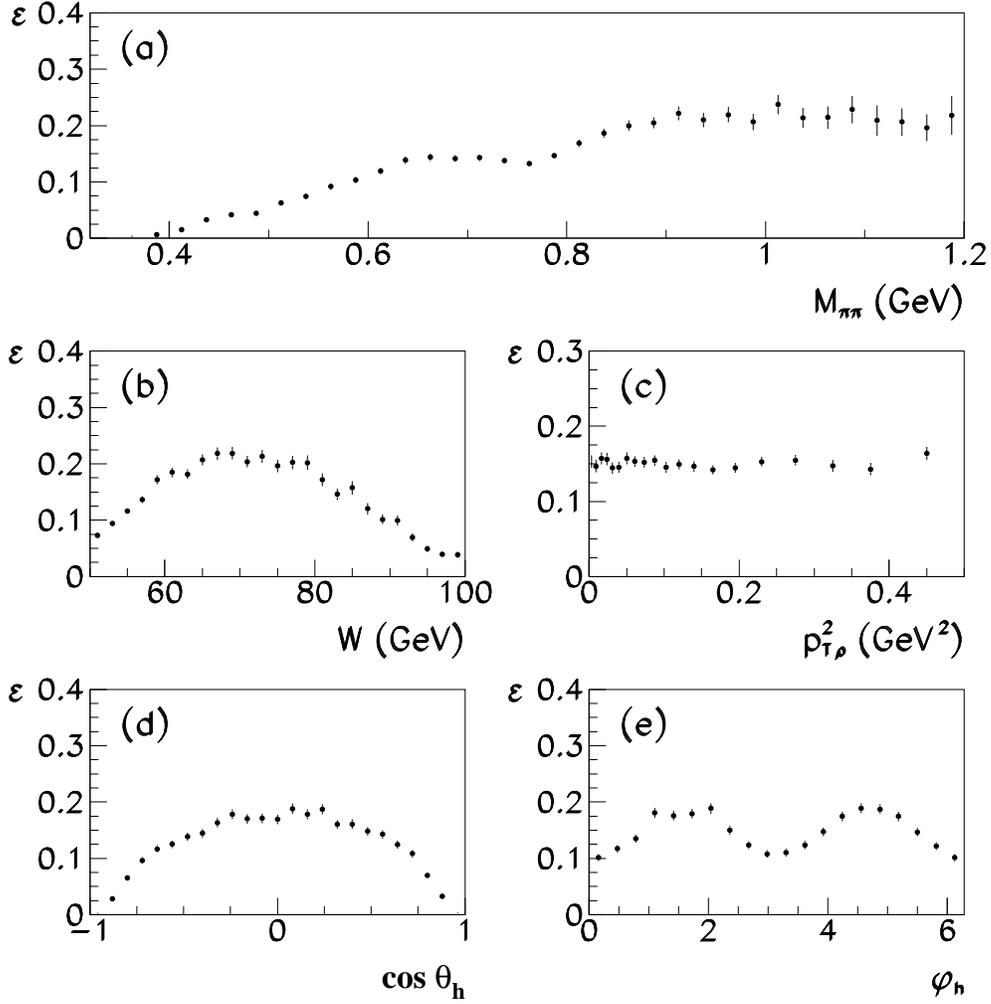}}
\end{center}
\vspace{1cm}
\caption{Overall acceptance $\varepsilon$ for elastic $\rho^0$ photoproduction,
$e p \rightarrow e \rho^0 p$, 
as a function of (a)~$M_{\pi\pi}$, (b)~$W$, (c)~$p_{T\rho}^2$, 
(d)~$\cos{\theta_h}$
and (e)~$\varphi_h$ obtained with the EPSOFT generator. 
Only statistical errors are shown. 
}
\label{fig_acceptance}
\end{figure}

\begin{figure}
\vspace{-2.4cm}
\begin{center}
\leavevmode
\hbox{%
\epsfxsize = 15cm
\epsffile{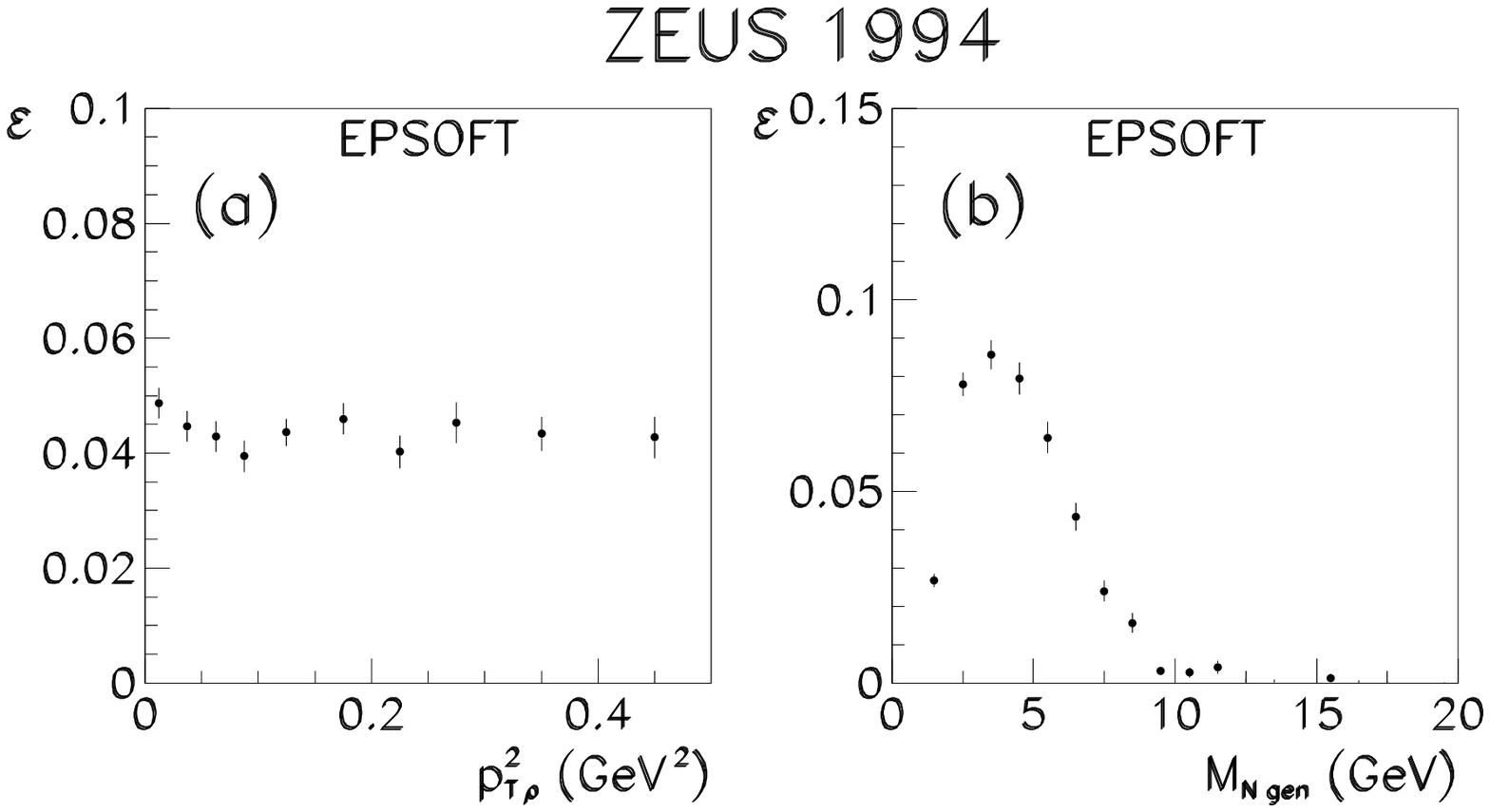}}
\end{center}
\vspace{1cm}
\caption{
The overall acceptance as a function of (a)~$p_{T\rho}^2$ and 
(b)~$M_{Ngen}$ for proton-dissociative events, $e p \rightarrow e \rho^0 N$,  
tagged with the PRT1 (see section~\protect\ref{selection_inelastic}); 
$M_{Ngen}$ indicates the generated value of $M_N$. 
Only statistical errors are shown. 
}
\label{fig_acceptance2}
\end{figure}

Figure~\ref{fig_acceptance} shows the overall acceptance
for elastic events
as a function of $M_{\pi\pi}$, $W$, $p_{T\rho}^2$, $\cos{\theta_h}$ and 
$\varphi_h$
obtained using EPSOFT. The average acceptance is 15\%. The dip in the 
acceptance at the $M_{\pi\pi}$ value corresponding to the $\rho^0$ peak 
is a consequence of the mass resolution.
The acceptance as a function of $p_{T\rho}^2$ and $M_N$, for 
proton-dissociative events tagged with the PRT1, is 
shown in Fig.~\ref{fig_acceptance2}. 
As for the elastic case, the acceptance is essentially independent of 
$p_{T\rho}^2$. While PYTHIA and EPSOFT give consistent results for the shape 
of the acceptance for the PRT1 tagged events (and for those tagged in the FCAL 
or in the LPS with $x_L<0.98$), the normalisation differs by up to a factor 
of two. 

\section{Elastic $\rho^0$ photoproduction}

\subsection{Background to elastic $\rho^0$ photoproduction}
\label{background_elastic}

After the selection cuts described in section~\ref{selection_elastic}, 
the main source of background was proton-dissociative events in which 
the mass $M_N$ was small and no particle from the system $N$ 
was detected. 

\begin{figure}
\vspace{-2.4cm}
\begin{center}
\leavevmode
\hbox{%
\epsfxsize = 15cm
\epsffile{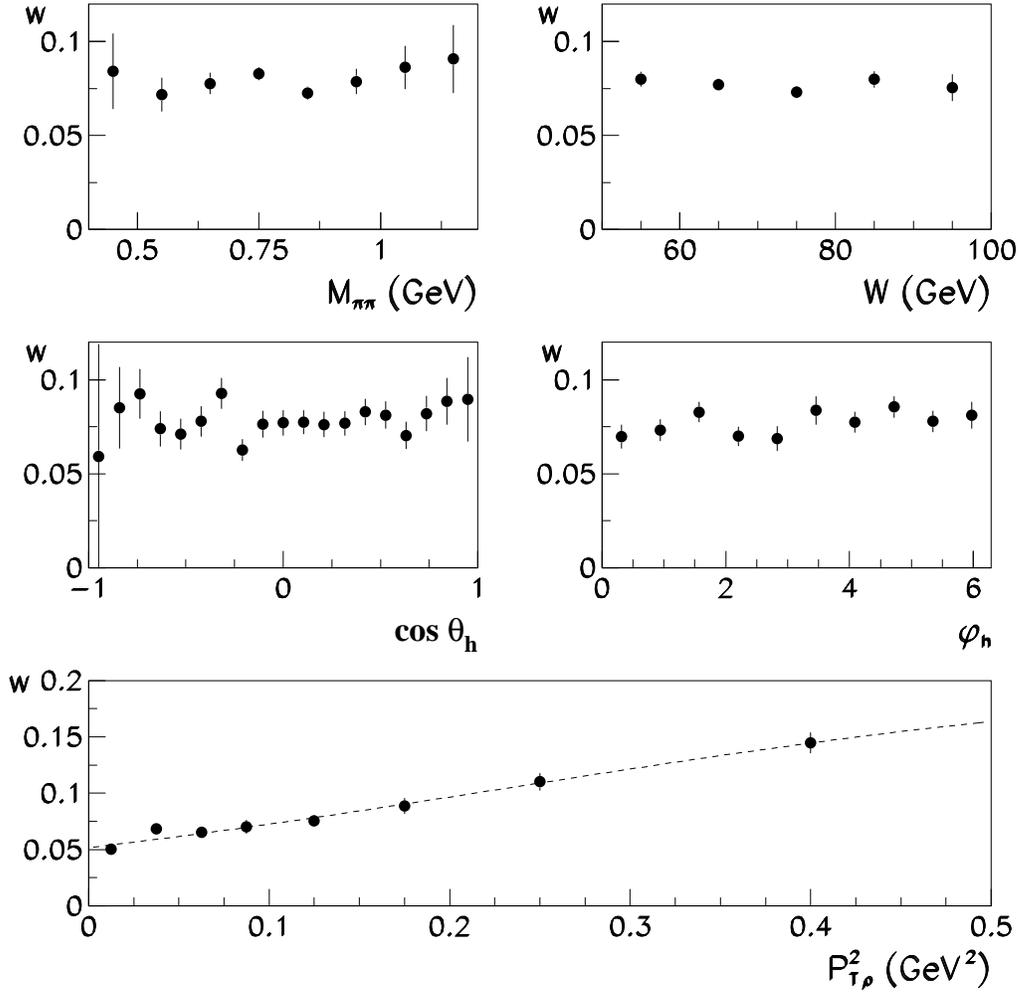}}
\end{center}
\vspace{1cm}
\caption{The ratios $w$
of the uncorrected $M_{\pi\pi}$, $W$, $\cos{\theta_h}$, $\varphi_h$ and
$p_{T\rho}^2$
distributions for the proton-dissociative sample (PRT tagged) and the sample 
obtained with the elastic cuts. Only statistical errors are shown.
The dashed line is the result of the fit with equation~(\protect\ref{diss}).
}
\label{ratio_pdiss} 
\end{figure}

The fraction of proton-dissociative events in the sample selected with the cuts 
of section~\ref{selection_elastic} 
was determined as follows. Proton-dissociative events were selected with the 
PRT1 (or the FCAL, but we shall concentrate on the PRT1 tagged sample in the following)
as described in section~\ref{selection_inelastic}.
The ratios $w$
of the uncorrected $M_{\pi\pi}$, $W$, $\cos{\theta_h}$ and $\varphi_h$ 
distributions for the proton-dissociative sample (selected with the PRT1) 
and the sample 
obtained with the elastic cuts (for the period in which the PRT1 was 
operational) were found consistent with being flat, as
shown in Fig.~\ref{ratio_pdiss}.
Since according to both PYTHIA and EPSOFT the 
requirement of activity in the PRT1 (or the FCAL) does not affect the 
shape of the acceptance, this result indicates 
that proton-dissociative and elastic $\rho^0$ photoproduction have the same 
$M_{\pi\pi}$, $W$, $\cos{\theta_h}$ and $\varphi_h$ distributions.
On the contrary, the ratio of the $p_{T\rho}^2$ distributions, also
shown in Fig.~\ref{ratio_pdiss}, rises with $p_{T\rho}^2$; since for both 
reactions the acceptance has the same shape as a function of $p_{T\rho}^2$ (cf. 
Figs.~\ref{fig_acceptance} and~\ref{fig_acceptance2}), this indicates a 
shallower $p_{T\rho}^2$ dependence for the proton-dissociative events. 

The fraction of proton-dissociative events in the total sample was thus 
taken to depend on $p_{T\rho}^2$ only. The $p_{T\rho}^2$ dependence 
of the background was determined as follows. 
Let the $p_{T\rho}^2$ distribution for the proton-dissociative sample be 
parametrised as 
$dN_{diss}/dp_{T\rho}^2= A_{diss}\exp{(-b_{diss}^{app}p_{T\rho}^2)}$ and that
for the elastic sample as
$dN_{el}/dp_{T\rho}^2= A_{el}\exp{(-b^{app}p_{T\rho}^2)}$. Also, let 
$dN_{PRT-tag}/dp_{T\rho}^2$ and $dN_{el-cuts}/dp_{T\rho}^2$ indicate the
measured $p_{T\rho}^2$ distributions for the proton-dissociative sample 
selected with the PRT1 and for the sample obtained with the elastic cuts, 
respectively.  Then the ratio $w(p_{T\rho}^2)$ can be written as:

\begin{eqnarray}
w(p_{T\rho}^2)&=&\frac{dN_{PRT-tag}/dp_{T\rho}^2}{dN_{el-cuts}/dp_{T\rho}^2}\nonumber \\
&=&\frac{\varepsilon_{diss} dN_{diss}/dp_{T\rho}^2}{\varepsilon_{el}dN_{el}/dp_{T\rho}^2+\varepsilon_{diss}^{el-cuts}dN_{diss}/dp_{T\rho}^2}\nonumber \\
&=&\frac{\varepsilon_{diss}}{\varepsilon_{el}} \frac{dN_{diss}/dp_{T\rho}^2}{dN_{el}/dp_{T\rho}^2+(\varepsilon_{diss}^{el-cuts}/\varepsilon_{el})dN_{diss}/dp_{T\rho}^2}\nonumber \\
&\propto& \frac{dN_{diss}/dp_{T\rho}^2}{dN_{el}/dp_{T\rho}^2+f_{diss}dN_{diss}/dp_{T\rho}^2} \label{diss0}\\ 
&=& \frac{A_{diss}\exp{(-b_{diss}^{app}p_{T\rho}^2)}}
{A_{el}\exp{(-b^{app}p_{T\rho}^2)}+f_{diss} A_{diss}\exp{(-b_{diss}^{app}p_{T\rho}^2)}} \nonumber\\ 
&=&\frac{1/f_{diss}}{(A_{el}/f_{diss}A_{diss})\exp{[-(b^{app}-b_{diss}^{app})p_{T\rho}^2]}+1},
\label{diss}  
\end{eqnarray}

\noindent
where $\varepsilon_{diss}$ indicates the acceptance for proton-dissociative
events to pass the proton-dissociative cuts of section~\ref{selection_inelastic},
$\varepsilon_{el}$ indicates the acceptance for elastic events to pass the 
elastic cuts of section~\ref{selection_elastic}, and 
$\varepsilon_{diss}^{el-cuts}$ indicates the acceptance for 
proton-dissociative events to pass the 
elastic cuts. We introduced the proportionality symbol in eq.~(\ref{diss0}) 
to account for the $p_{T\rho}^2$-independent ratio of the acceptance for the
proton-dissociative events tagged by the PRT1 and that for the elastic 
events satisfying the
elastic cuts. The quantity $f_{diss}$ is the ratio of the acceptance for 
proton-dissociative events passing the elastic cuts and that for
elastic events passing the elastic cuts; this ratio is taken to be
$p_{T\rho}^2$-independent.

The difference $(b^{app}-b_{diss}^{app})$ was determined by fitting eq.~(\ref{diss})
to the data in the range $0<p_{T\rho}^2<0.5$~GeV$^2$ and was found to be 
$4.8 \pm 1.5~\mbox{(stat.)} \pm 0.5~\mbox{(syst.)}$~GeV$^{-2}$ for the 
proton-dissociative events tagged with the PRT1. The result of the fit
is shown as the dashed line in Fig.~\ref{ratio_pdiss}.
The systematic uncertainty reflects the sensitivity 
of the result to the limits of the fitted range, with the lower limit 
varied between $p_{T\rho}^2= 0$ and 0.075~GeV$^{2}$ and the upper one 
between 0.3 and 0.5~ GeV$^{2}$.
The proton-dissociative events tagged with the FCAL yield 
$(b^{app}-b_{diss}^{app}) =4.1 \pm 2.0~\mbox{(stat.)}$~GeV$^{-2}$.

To determine the normalisation of the proton-dissociative background the following 
procedure was adopted. As discussed earlier (cf. section~\ref{sec_montecarlo})
the acceptance for the proton-dissociative events depends on the Monte Carlo 
program; hence the proton-dissociative sample was not used. 
Instead we used the sample 
satisfying the elastic cuts and its subsample~\cite{lps_rho,thesis_dirk} in 
which the presence of a high momentum ($x_L >$ 0.98) proton in the LPS 
identified elastic events. 
The region $0.075<p_{T\rho}^2<0.5$~GeV$^2$ was 
used, where the acceptance of the LPS is well understood~\cite{lps_rho}. 

The function $r(p_{T\rho}^2)$ was introduced, defined as the fraction 
of proton-dissociative events in the elastic sample:

\begin{eqnarray}
r(p_{T\rho}^2)=\frac{ dN_{el-cuts}/dp_{T\rho}^2 - dN_{LPS}/dp_{T\rho}^2}{dN_{el-cuts}/dp_{T\rho}^2},
\label{diss01}  
\end{eqnarray}
\noindent
where $dN_{LPS}/dp_{T\rho}^2$ is the yield, corrected for the LPS acceptance, 
for the LPS tagged elastic events ($x_L>0.98$). Using the notation introduced
earlier, $r(p_{T\rho}^2)$ can be rewritten as 

\begin{eqnarray}
r(p_{T\rho}^2)=\frac{1}{D\exp{[-(b^{app}-b_{diss}^{app}) p_{T\rho}^2]} +1},
\label{diss1}  
\end{eqnarray}
\noindent
where $(b^{app}-b_{diss}^{app})$ was taken to be 
$4.8 \pm 1.5~\mbox{(stat.)} \pm 0.5~\mbox{(syst.)}$~GeV$^{-2}$, 
as discussed above, and a fit to the data gave 
$D=7.3^{+1.2}_{-0.9}$~(stat.)~$^{+3.2} _{-2.1}$~(syst.).
In order to correct for the proton-dissociative background each event was then 
weighted by $[1-r(p_{T\rho}^2)]$. 
The resulting integrated fraction of proton-dissociative background in 
the untagged sample is 
$R_{diss}= (20 \pm~2~\mbox{(stat.)}~^{+6}_{-5}~\mbox{(syst.)})\%$
for $p_{T\rho}^2<0.5$~GeV$^2$.

In summary, function~(\ref{diss}) was fitted to proton-dissociative events 
tagged with the PRT1 to determine $(b^{app}-b_{diss}^{app})$. The latter
was then used in function~(\ref{diss1}), which was evaluated using
the yields for purely elastic events tagged by the LPS ($x_L>0.98$) 
and for the events passing the elastic cuts.
The fit result was used to evaluate the normalisation of the proton-dissociative
contamination in the sample selected with the elastic cuts 
and hence the overall contamination
$R_{diss}$.

\bigskip
Positron beam-gas and proton beam-gas contaminations were estimated 
from the unpaired bunch event samples to which the selection criteria
described earlier were applied.
The number of events passing the cuts was then scaled by the 
ratio between the positron (proton) current in the paired bunches and
the current in the positron (proton) unpaired bunches. The contamination 
due to positron-gas interactions was estimated to be (0.6 $\pm$ 0.1)\%,
while that due to proton-gas events was found to be 
$\ltap (0.01 \pm 0.01)\%$. 
The contamination from elastic production of $\omega$ and $\phi$ mesons 
(decaying to $\pi^+ \pi^- \pi^0$) was estimated by using simulated 
events and found to be $\lsim 0.5\%$~\cite{thesis_dirk}.

All subsequent results are shown after subtraction of 
the contributions from proton-dissociative events, beam-gas 
interactions, $\omega$ and $\phi$ production.

\subsection{Results for elastic $\rho^0$ photoproduction}
\label{results_elastic}

\subsubsection{Cross section determination}

The differential and integrated photoproduction cross sections for 
the process $\gamma p \rightarrow \pi^+ \pi^- p$ were obtained from the 
event yield measured for the reaction 
$ep \rightarrow e \pi^+ \pi^- p$. The cross sections for these two 
processes are related by

\begin{eqnarray}
\frac{d^2\sigma_{ep \rightarrow e\pi^+\pi^-p}}{dydQ^2} &=& \frac{\alpha}{2\pi Q^2}
\left[\left( \frac{1+(1-y)^2}{y} - \frac{2(1-y)}{y}\cdot \frac{Q_{\min}^2}{Q^2}\right) \cdot
\sigma_T^{\gamma^*p \rightarrow \pi^+\pi^- p}(W,Q^2) \right. \nonumber \\
& &\left. + \frac{2(1-y)}{y} \cdot
\sigma_L^{\gamma^*p \rightarrow \pi^+\pi^-p}(W,Q^2)\right],
\label{bornc}
\end{eqnarray}
where $\alpha$ is the fine structure constant and
$\sigma_T^{\gamma^*p \rightarrow \pi^+\pi^-p}(W,Q^2)$ and
$\sigma_L^{\gamma^*p\rightarrow \pi^+\pi^-p}(W,Q^2)$
are the cross sections for transversely and longitudinally
polarised virtual photons, respectively. These cross sections are assumed 
to be

\begin{eqnarray}
\sigma_T^{\gamma^*p \rightarrow \pi^+\pi^-p}(W,Q^2) =
\sigma_{\gamma p\rightarrow \pi^+\pi^-p}(W) \left/
\left(1+\frac{Q^2}{M_{\rho}^2}\right)^2 \right.,
\label{sigmat}
\end{eqnarray}

\noindent
for transversely polarised photons and 

\begin{eqnarray}
\sigma_L^{\gamma^*p \rightarrow \pi^+\pi^-p}(W,Q^2) =
\sigma_T^{\gamma^*p \rightarrow \pi^+\pi^-p}(W,Q^2)
\cdot \frac{Q^2}{M_{\rho}^2} \xi^2
\label{sigmal}
\end{eqnarray}

\noindent
for longitudinally polarised photons, where $\xi$ is a proportionality
constant of order unity (cf. e.g.~\cite{bauer}).
The results presented in this paper are insensitive to the value of $\xi^2$:
varying $\xi^2$ between 0 and 1 has negligible effects.

Substituting the latter two expressions into equation~(\ref{bornc})
yields:
\begin{eqnarray}
\frac{d^2\sigma_{ep \rightarrow e\pi^+\pi^-p }}{dy dQ^2}
=\varphi(y,Q^2) \cdot \sigma_{\gamma p \rightarrow \pi^+\pi^-p }(W(y)),
\label{crs}
\end{eqnarray}
  
\noindent
which defines the effective photon flux $\varphi(y,Q^2)$.

From eq.~(\ref{crs}), knowing the effective photon flux, it was then possible to
determine the cross section
$\sigma_{\gamma p \rightarrow \pi^+\pi^-p }$. As an example,
the differential cross section 
$d\sigma_{\gamma p \rightarrow \pi^+ \pi^- p}/dM_{\pi \pi}$ 
was evaluated in each  
$M_{\pi \pi}$ bin of width $\Delta M_{\pi \pi}$ as 

\begin{eqnarray}
\frac{d\sigma_{\gamma p \rightarrow \pi^+ \pi^- p}}{dM_{\pi\pi}}=\frac {N_{\pi^+\pi^-}}{A \cdot L \cdot \Phi \cdot \Delta M_{\pi \pi} },
\label{crosssection}
\end{eqnarray}

\noindent
where $N_{\pi^+\pi^-}$ is the number of observed events in the bin after 
background subtraction and correction for the RCAL trigger efficiency, 
$L$ the integrated luminosity and $A$ the overall acceptance in 
the bin excluding the RCAL efficiency. The integral of the 
effective photon flux $\varphi(y,Q^2)$ over the $y$ and $Q^2$ ranges
covered by the experiment is indicated as $\Phi$. In the following, for
brevity, the subscript $\gamma p \rightarrow \pi^+ \pi^- p$ will be dropped.

\subsubsection{Differential cross section 
$d\sigma/dM_{\pi\pi}$ for the reaction $\gamma p \rightarrow \pi^+ \pi^- p$}
\label{mass}

\begin{figure}
\vspace{-4cm}
\begin{center}
\leavevmode
\hbox{%
\epsfxsize = 17cm
\epsffile{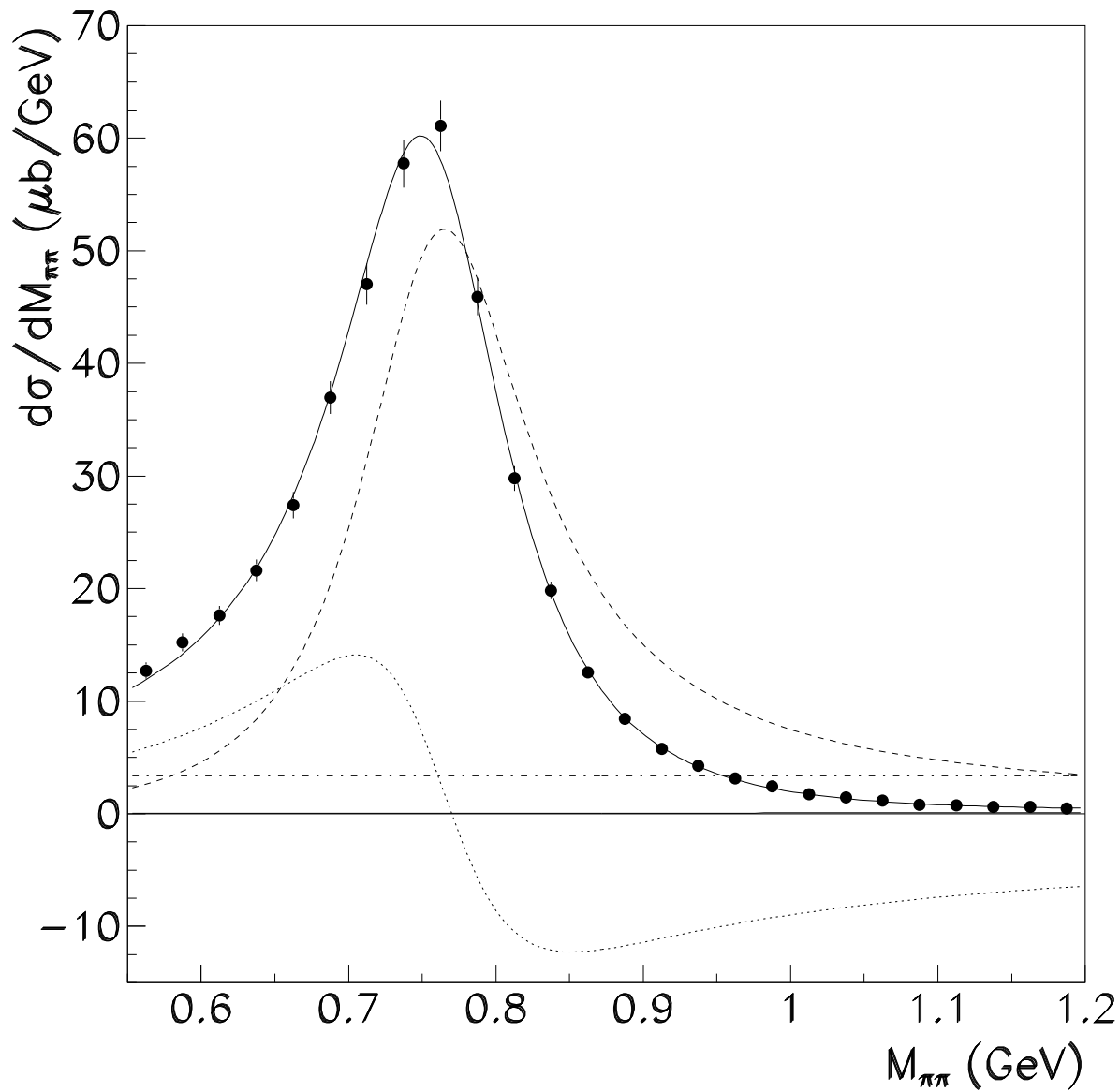}}
\end{center}
\vspace{1cm}
\caption{The differential cross section $d\sigma/dM_{\pi\pi}$ for the
elastic reaction $\gamma p \rightarrow  \pi^+ \pi^- p$ in the kinematic region
$50<W<100$~GeV and $|t|<0.5$~GeV$^2$. The points represent 
the ZEUS data and the curves indicate the result of the fit to the data
using expression~(\protect\ref{equsod}).
The dashed curve 
represents the resonant contribution, the dot-dashed curve the non-resonant 
contribution and the dotted curve the contribution of the interference term.
The continuous curve is the sum. 
Only statistical errors are shown. 
}
\label{fig_mass_tot} 
\end{figure}

\begin{figure}
\vspace{-4cm}
\begin{center}
\leavevmode
\hbox{%
\epsfxsize = 17cm
\epsffile{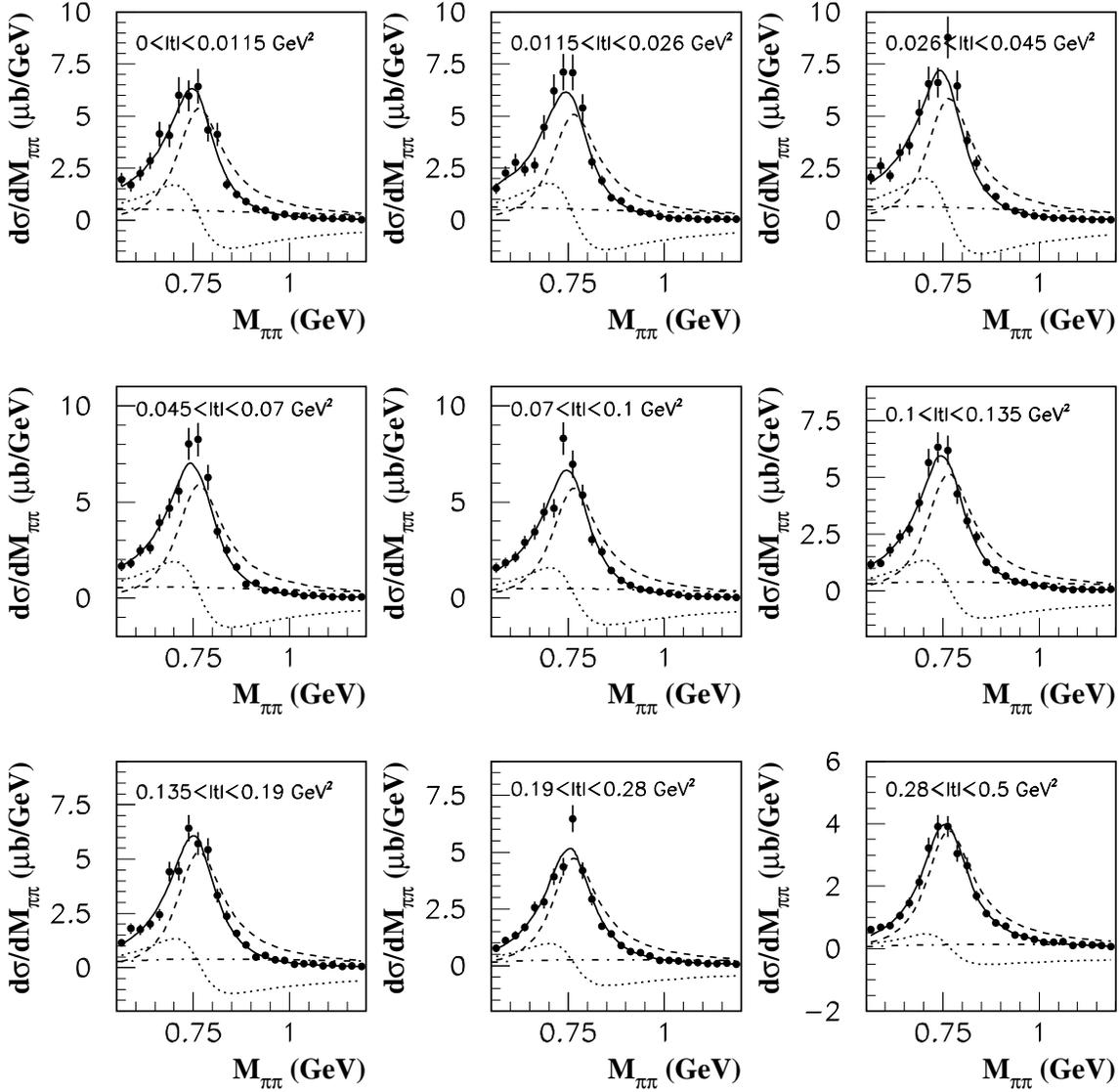}}
\end{center}
\vspace{1cm}
\caption{The differential cross sections $d\sigma/dM_{\pi\pi}$ 
for the
elastic reaction $\gamma p \rightarrow  \pi^+ \pi^- p$ in the range 
$50<W<100$~GeV 
for different $|t|$ bins. The points represent 
the ZEUS data and the curves indicate the results of the fit 
to the data based on the model of ref.~\protect\cite{misha_yuly} 
(cf. section~\protect\ref{sec_pi_p}). The dashed curve 
represents the resonant contribution, the dot-dashed curve the non-resonant 
contribution and the dotted curve the contribution of the interference term.
The continuous curve is the sum.
Only statistical errors are shown.}
\label{fig_mass_bin} 
\end{figure}

In Fig.~\ref{fig_mass_tot} the differential cross section 
$d\sigma/dM_{\pi\pi}$ for the process $\gamma p \rightarrow \pi^+ \pi^- p$
is shown in the kinematic range $|t|<0.5$~GeV$^2$ and $50<W<100$~GeV.
Figure~\ref{fig_mass_bin} shows $d\sigma/dM_{\pi\pi}$ for 
different $t$ bins. 
The $\rho^0$ resonance shape is skewed, as observed 
in previous measurements~\cite{bauer}-\cite{lps_rho}. This can be 
understood in terms of the interference between resonant 
$\rho^0$ production and non-resonant $\pi^+\pi^-$ 
production~\cite{soeding,misha_yuly}. 

Fits to the points of Fig.~\ref{fig_mass_tot} 
were performed in the range $0.55 < M_{\pi\pi} < 1.2$ GeV 
with the following parametrisation~\cite{rho93}:
\begin{eqnarray}
\frac{d\sigma}{dM_{\pi\pi}} = 
\left | A\frac{ \sqrt{M_{\pi\pi} M_\rho \Gamma_\rho}}{M^2_{\pi\pi}-
M_\rho^2+iM_\rho\Gamma_\rho}+B \right | ^2 + f_{PS},
\label{equsod}
\end{eqnarray}
where $M_{\rho}$ is the nominal $\rho^0$ mass and 
$\Gamma_\rho=\Gamma_0(p^*/p_0^*)^3 (M_{\rho}/M_{\pi\pi})$ the momentum 
dependent $\rho^0$ width, with $\Gamma_0$ the $\rho^0$ width, $p^*$ the $\pi$ 
momentum in the $\pi\pi$ rest frame and $p_0^*$ the value of $p^*$ at 
the $\rho^0$ nominal mass. 
The non-resonant amplitude is denoted by $B$ and is assumed to be constant 
and real. The term $f_{PS}$ is a first order polynomial of the form 
$f_{PS}=A_{PS}(1+B_{PS}M_{\pi\pi}$)
accounting for residual background from the process 
$\gamma p \rightarrow Xp$. The coefficient $B_{PS}$ was
estimated to be $B_{PS}=1.5$~GeV$^{-1}$ from Monte Carlo studies in which
PYTHIA~\cite{pythia} was used to simulate the reaction $ep \rightarrow eXp$.
The fitted value of $A_{PS}$ corresponds to an integrated 
contribution typically smaller than 1\% of the total, independent of $W$ and 
$p_T^2$. 
Table~\ref{tab_mass_fit} gives 
the parameters of the fit for the spectrum shown in 
Fig.~\ref{fig_mass_tot}. The fitted values of the $\rho^0$ mass and
width are consistent with the Particle Data Group tables~\cite{pdg} and
the $\chi^2/ndf$ for the fit is 15.3/21. 

The fits to $d\sigma/dM_{\pi\pi}$ were repeated in a different way. 
The mass spectrum was corrected for the acceptance excluding the effects of 
migration. The mass spectrum was 
then fitted with formula~(\ref{equsod}) convoluted with a Gaussian which 
describes the detector resolution. The width of the Gaussian varied 
between 6 and 14~MeV depending on $W$.
The resulting values for the $\rho^0$ mass and width
were $771 \pm 2$~MeV and $159 \pm 3$~MeV. The difference between this value
of the width and that given in Table~\ref{tab_mass_fit} gives an indication
of the systematic errors associated to the fit.
The other parameters 
of the fit were found to be $A=-2.75 \pm 0.04$~$\mu$b$^{1/2}$,
$B=1.94\pm 0.07$~$~\mu$b$^{1/2}$ GeV$^{-1/2}$ and 
$A_{PS}=0.000 \pm 0.015$~$\mu$b/GeV. The results for the cross section
do not change if this method is used. A fit including a $\rho^0$-$\omega$
interference term was also performed~\cite{thesis_dirk}; this led to a 
slightly better $\chi^2$. However, none of the results presented in the following
changes if such a fit is used.

\begin{table}
\begin{center}
\begin{tabular}{lcc} \hline \hline
Parameter           & value            & stat. error       \\ \hline
$M_{\rho}$          & 0.770     GeV        &0.002        GeV         \\
$\Gamma_0$          & 0.146     GeV        &0.003        GeV         \\
$A$                 &$-2.75~\mu$b$^{1/2}$ &0.04~$\mu$b$^{1/2}$      \\
$B$                 &$1.84~\mu$b$^{1/2}$ GeV$^{-1/2}$&0.06~$\mu$b$^{1/2}$ GeV$^{-1/2}$  \\ 
$A_{PS}$            &$0.030~\mu$b/GeV        & 0.015~$\mu$b/GeV\\ \hline \hline
\end{tabular}
\end{center}
\caption{Results of the fit to the mass spectrum 
of Fig.~\protect\ref{fig_mass_tot} for $50~<W<100$~GeV and $|t|<0.5$~GeV$^2$ 
using expression~(\protect\ref{equsod}). Only statistical errors are given.}
\label{tab_mass_fit}
\end{table}

The curves shown in Fig.~\ref{fig_mass_bin} were obtained using a
calculation~\cite{misha_yuly} based
on S\"oding's model~\cite{soeding}. In this case the mass and the width 
of the $\rho^0$ were fixed to the values given in the Particle Data 
Group tables~\cite{pdg}. The $\pi$-$p$ total cross section, 
a free parameter of model~\cite{misha_yuly}, was fitted.
This fit is discussed in detail in section~\ref{sec_pi_p}; here we only 
remark that the results of the calculation are in good agreement with 
the data with an average $\chi^2/ndf$ of 1.0. The 
non-resonant and interference terms are also shown in the figure; 
the non-resonant contribution varies very little
with $M_{\pi \pi}$, a result consistent with the ansatz, made above,
that $B$ is a constant, as assumed in our previous 
analyses~\cite{rho93,lps_rho}.

\begin{figure}
\vspace{-4cm}
\begin{center}
\leavevmode
\hbox{%
\epsfxsize = 17cm
\epsffile{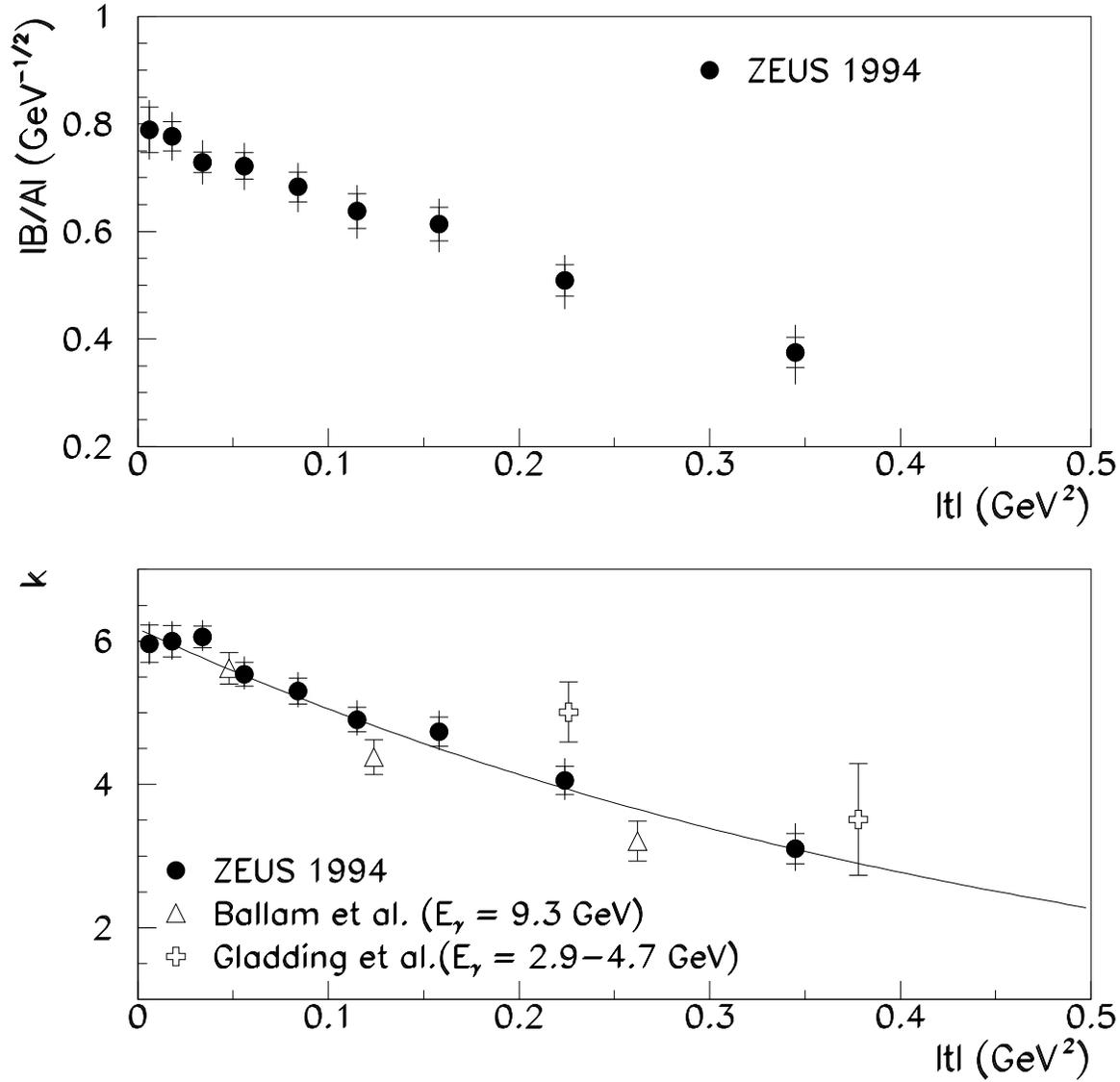}}
\end{center}
\vspace{1cm}
\caption{The ratio $|B/A|$ (upper plot) and the parameter $k$ (lower plot) 
as a function of $|t|$ obtained by fitting 
eq.~(\protect\ref{equsod}) and eq.~(\protect\ref{equross}), respectively, to 
the points of Fig.~\protect\ref{fig_mass_bin} for the 
elastic reaction $\gamma p \rightarrow  \pi^+ \pi^- p$ in the region
$50<W<100$~GeV. The solid points represent the ZEUS measurements. 
The inner error bars 
indicate the statistical uncertainty, the outer ones the statistical
and systematic uncertainties added in quadrature. The results of the fixed
target experiments~\protect\cite{ballam} and ~\protect\cite{gladding} 
(labelled ``Ballam et al." and ``Gladding et al.", respectively) are also shown. 
The continuous line indicates the effective expectation of the
S\"oding model as implemented in the 
calculation by Ryskin and Shabelski~\protect\cite{misha_yuly} 
(cf. section~\protect\ref{sec_pi_p}). 
}
\label{fig_basod}
\end{figure}

Fits to the data of Fig.~\ref{fig_mass_bin} using 
formula~(\ref{equsod}) were also carried out, with the mass and the width of 
the $\rho^0$ fixed to the values given in table~\ref{tab_mass_fit}. 
The results for $|B/A|$ from these fits are shown as a function of $|t|$ in
the upper plot of Fig.~\ref{fig_basod}:  $|B/A|$ decreases with increasing
$|t|$. The quantity $|B/A|$ is a measure of the ratio of the non-resonant to 
resonant contributions; its decrease with increasing 
$|t|$ was already observed in fixed 
target experiments~\cite{bauer} and can be described in the framework 
of the S\"oding model~\cite{soeding,misha_yuly}. 

Alternatively, the following phenomenological parametrisation 
proposed by Ross and Stodolsky~\cite{rosssto} was used to fit 
the mass distribution:
\begin{eqnarray}
\frac{d\sigma}{dM_{\pi\pi}} = f_{\rho} \cdot BW_{\rho}(M_{\pi\pi}) \cdot 
(M_{\rho}/M_{\pi\pi})^k + f_{PS},
\label{equross}
\end{eqnarray}
where $BW_{\rho}(M_{\pi\pi})$ is a relativistic $p$-wave Breit-Wigner function 
and the factor $(M_{\rho}/M_{\pi\pi})^k$ accounts for the skewing of 
the signal. In this case the fitted values of the $\rho^0$ mass and width
are $771\pm2$~MeV and $138\pm3$~MeV, respectively; the parameter $k$ 
is $5.13 \pm 0.13$. Here again the fits were repeated in different $|t|$ bins,
keeping the mass and the width of the $\rho^0$ fixed to the values just quoted. 
The parameter $k$ is plotted as a function of $|t|$ 
in the lower part of Fig.~\ref{fig_basod}. 
The decrease of the amount 
of skewing with increasing $|t|$ is, in this case, reflected in the decrease of 
$k$. Our results are in agreement with those
found in fixed target photoproduction experiments~\cite{ballam,gladding},
indicating that skewing of the $\rho^0$ resonant shape depends only weakly,
if at all, on $W$. Note that in all $t$ bins the median $Q^2$ is 
lower than $10^{-5}$~GeV$^{2}$. The results are 
consistent with the effective expectation of the 
S\"oding model~\cite{soeding,misha_yuly}, as the 
continuous line in the lower plot of Fig.~\ref{fig_basod} shows (cf. section~\ref{sec_pi_p}).

\begin{figure}
\vspace{-4cm}
\begin{center}
\leavevmode
\hbox{%
\epsfxsize = 17cm
\epsffile{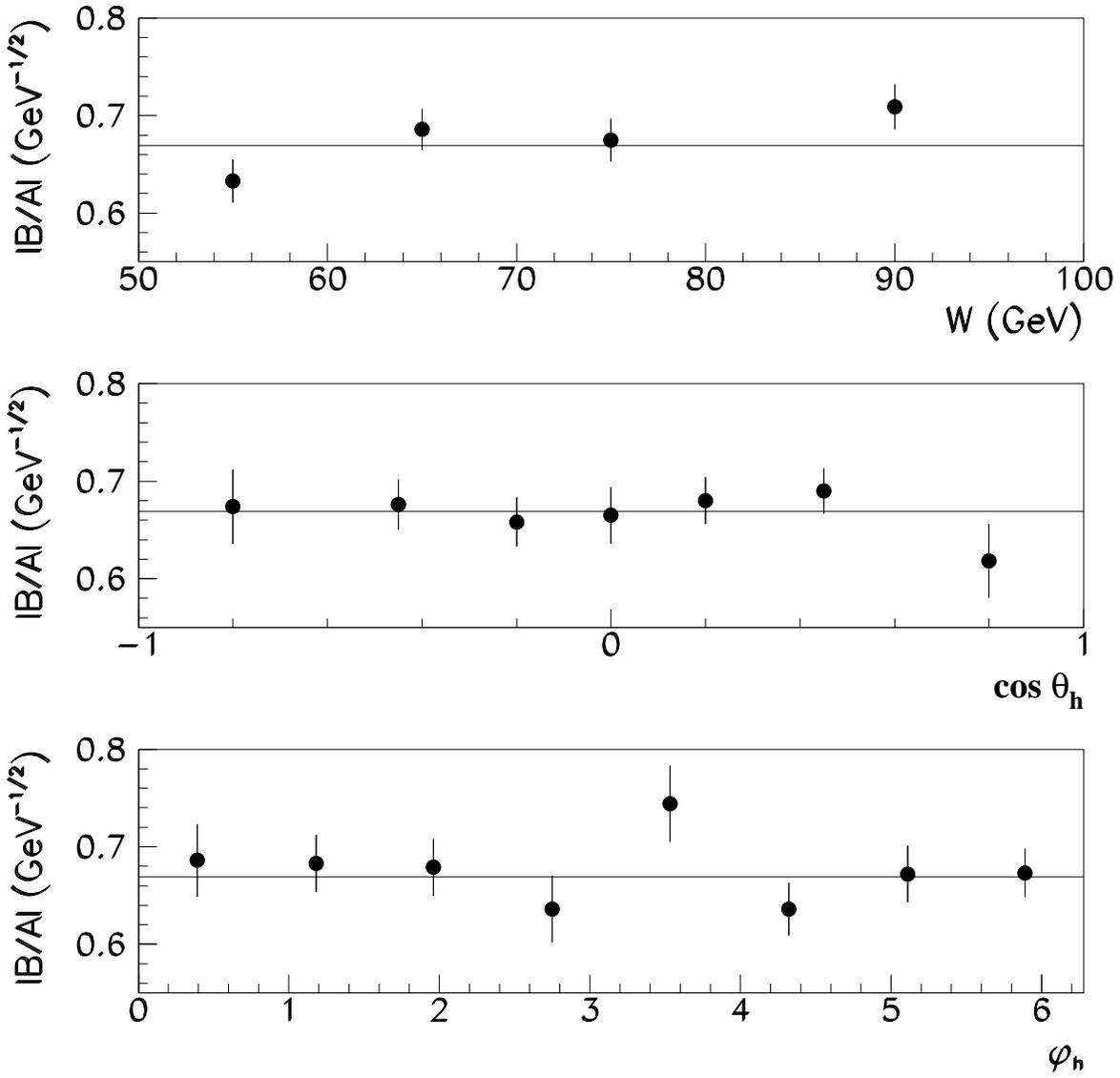}}
\end{center}
\vspace{1cm}
\caption{The ratio $|B/A|$ as a function of $W$, $\cos{\theta_h}$ and 
$\varphi_h$ for the elastic reaction $\gamma p \rightarrow  \pi^+ \pi^- p$
in the kinematic range $0.55<M_{\pi\pi}<1.2$~GeV, $50 <W<100$~GeV and 
$|t|<0.5$~GeV$^2$.
Statistical errors only are shown. The continuous lines indicate 
the average value of $|B/A|$.
}
\label{fig_basod_1}
\end{figure}

Fits using formula~(\ref{equsod}) were also performed in bins of $W$,
$\cos{\theta_h}$ and $\varphi_h$, again with the mass and the width of 
the $\rho^0$ fixed to the values given in table~\ref{tab_mass_fit}. 
The ratios $|B/A|$ from these fits are shown in
Fig.~\ref{fig_basod_1};  $|B/A|$ appears to be independent of $W$ 
(as already suggested by the comparison with 
the fixed target data, cf. Fig.~\ref{fig_basod}) as well as
of the decay pion polar and azimuthal angles in the 
helicity frame.

\subsubsection{Integrated $\gamma p \rightarrow \rho^0 p$ cross section}
\label{integrated}

\begin{figure}
\vspace{-4.0cm}
\begin{center}
\leavevmode
\hbox{%
\epsfxsize = 15cm
\epsffile{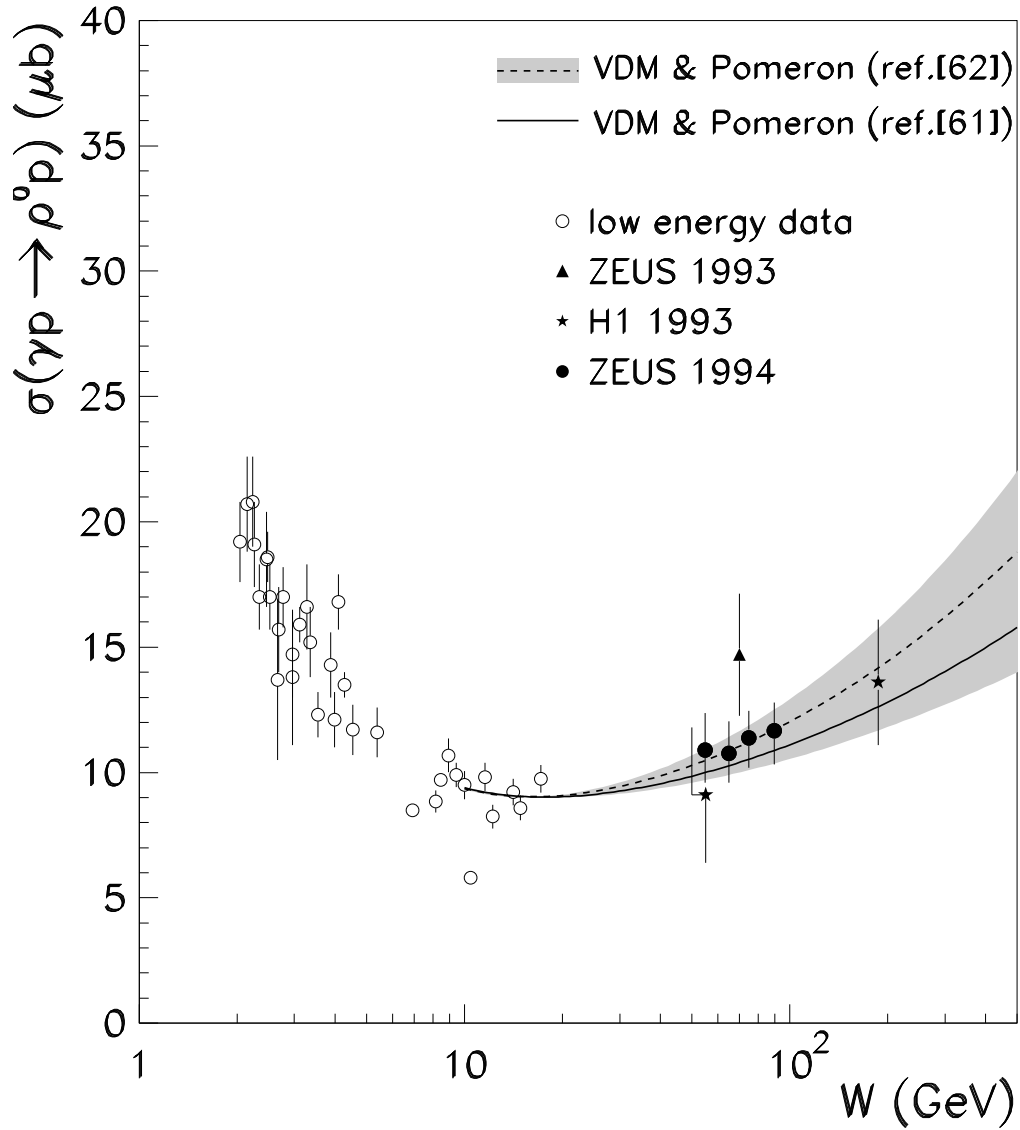}}
\end{center}
\vspace{1cm}
\caption{The integrated cross section 
$\sigma_{\gamma p \rightarrow \rho^0 p}$
as a function of the centre-of-mass energy $W$. The ZEUS results are 
given for the range $2M_{\pi}<M_{\pi\pi}<M_{\rho}+5\Gamma_0$, 
$|t|<0.5$~GeV$^2$. The other results from 
HERA~\protect\cite{rho93,h1rho} and a compilation of low energy 
data~\protect\cite{berger}-\protect\cite{gladding},  
~\protect\cite{struczinski}-\protect\cite{omega} 
are also shown. 
The continuous and dashed line are parametrisations~\protect\cite{thesis_joern}
based on Regge theory 
which assume the value of the pomeron intercept found by 
Donnachie and Landshoff~\protect\cite{dola} and by 
Cudell et al.~\protect\cite{Cudell}, respectively. 
The band corresponds to the uncertainty
in the determination of the pomeron intercept of ref.~\protect\cite{Cudell}. 
The error bars of the ZEUS points indicate the sum of 
statistical and systematic uncertainties in quadrature. For the points 
at the same value of $W$, the error bars have been offset.
}
\label{fig_cross}
\end{figure}

The integrated $\gamma p \rightarrow \rho^0 p$ cross section for 
$|t|< 0.5$ ${\rm GeV}^2$ was determined in four $W$ bins. 
In each of these bins fits to the mass spectra were performed using 
equation~(\ref{equsod}); $M_{\rho}$,
$\Gamma_{0}$ and $B/A$ were fixed to the values given in table~\ref{tab_mass_fit}. 
Following refs.~\cite{rho93,h1rho}, the cross 
section was calculated by integrating the resonant contribution obtained 
from the fit over the range $2M_\pi < M_{\pi \pi} < M_\rho + 5 \Gamma_0$.
Figure~\ref{fig_cross} and Table~\ref{tabcross} show the results. 
Table~\ref{tabcross_pi} gives the results for the reaction 
$\gamma p \rightarrow \pi^+\pi^- p$ over the same mass range; it was obtained
by integrating the result for the first term in eq.~(\ref{equross}).

\begin{table}[t]\centering
\begin{tabular}{cc}
\hline \hline
$\langle {\rm W} \rangle$ [GeV] & $\sigma_{\gamma p \to \rho^0 p}$ [$\mu$b] \\ 
\hline
 55 & 10.9 $\pm$ $0.2~(\mbox{stat.}) ^{+1.5}_{-1.3}~(\mbox{syst.}) $ \\ 
\hline
 65 & 10.8 $\pm$ $0.2~(\mbox{stat.}) ^{+1.3}_{-1.1}~(\mbox{syst.}) $ \\ 
\hline
 75 & 11.4 $\pm$ $0.3~(\mbox{stat.}) ^{+1.0}_{-1.2}~(\mbox{syst.}) $ \\ 
\hline
 90 & 11.7 $\pm$ $0.3~(\mbox{stat.}) ^{+1.1}_{-1.3}~(\mbox{syst.}) $ \\ 
\hline \hline
\end{tabular}
\caption{Elastic $\rho^0$ photoproduction cross section for 
$|t|< 0.5$ ${\rm GeV}^2$, integrated over the mass range 
$2M_\pi < M_{\pi \pi} < M_\rho + 5 \Gamma_0$ in four $W$ bins. 
The results were calculated by integrating the resonant contribution obtained 
from the fit with eq.~(\protect\ref{equsod}).}
\label{tabcross}
\end{table}

\begin{table}[t]\centering
\begin{tabular}{cc}
\hline \hline
$\langle {\rm W} \rangle$ [GeV] & $\sigma_{\gamma p \to \pi^+ \pi^- p}$ [$\mu$b] \\ 
\hline
 55 & 12.2 $\pm$ $0.2~(\mbox{stat.}) ^{+1.6}_{-1.4}~(\mbox{syst.}) $ \\ 
\hline
 65 & 12.1 $\pm$ $0.2~(\mbox{stat.}) ^{+1.2}_{-1.2}~(\mbox{syst.}) $ \\ 
\hline
 75 & 12.8 $\pm$ $0.3~(\mbox{stat.}) ^{+1.1}_{-1.3}~(\mbox{syst.}) $ \\ 
\hline
 90 & 13.1 $\pm$ $0.3~(\mbox{stat.}) ^{+1.2}_{-1.5}~(\mbox{syst.}) $ \\ 
\hline \hline
\end{tabular}
\caption{Elastic $\pi^+\pi^-$ photoproduction cross section for 
$|t|< 0.5$ ${\rm GeV}^2$, integrated over the mass range 
$2M_\pi < M_{\pi \pi} < M_\rho + 5 \Gamma_0$ in four $W$ bins. 
The results were obtained 
by integrating the first term in eq.~(\protect\ref{equross}).}
\label{tabcross_pi}
\end{table}

The systematic uncertainties are dominated by 
the uncertainties on the acceptance (4-10\%),
the proton-dissociative background ($8.5\%$) and
the number of $\rho^0$ signal events, which depends on the functional 
form chosen~\cite{thesis_dirk} to fit the mass spectrum (4\%); 
the parameters $M_{\rho}$, $\Gamma_0$ 
and $B/A$ were also varied within their statistical errors ($1.5\%$).
The uncertainty on the acceptance (4-10\%) is $W$ dependent and 
has two main contributions:
the calorimeter trigger efficiency near the 
threshold (4-10\%) and the sensitivity of the results 
to the cuts (4-2\%).

Table~\ref{syserr} summarises the contributions to the systematic 
uncertainty.
The total systematic uncertainty was obtained by summing all contributions 
in quadrature. 

\begin{table}
\begin{center}
\begin{tabular}{lc}                            \hline \hline
 Contribution from                            &  Uncertainty \\ \hline 
 Luminosity                                   &  1.5\% \\ 
 Acceptance: trigger efficiency               &  4-10\%   \\
 Acceptance: sensitivity to cuts              &  4-2\%   \\
 $p$-dissociative background subtraction      &  8.5\%   \\
 Background due to elastic $\omega$ and $\phi$ production &  1\%   \\
 Procedure to extract the resonant part of the cross section & 4.5\% \\
 Radiative corrections             &  2\%   \\
 \hline
 Total                             & 11-14\% \\ 
 \hline \hline
\end{tabular}
\end{center}
\caption{Individual and total contributions to the systematic uncertainty on the 
integrated cross section. 
}
\label{syserr}
\end{table}

Figure~\ref{fig_cross} includes a partial compilation of low energy 
measurements, as well as the recent ZEUS~\cite{rho93} and 
H1~\cite{h1rho} results. Also shown are parametrisations~\cite{thesis_joern} 
based on Regge theory
which assume the value of the pomeron intercept found by 
Donnachie and Landshoff~\cite{dola} and by Cudell et al.~\cite{Cudell},
respectively.
The $W$ dependence of the data is described satisfactorily by both.

A least squares fit to the present data alone with a function of the type 
$\sigma_{\gamma p \to \rho^0 p}(W)=\sigma_{\gamma p \to \rho^0 p}(W_0)(W/W_0)^a$ 
gives 
$\sigma_{\gamma p \to \rho^0 p}(W_0)= 
11.2\pm 0.1~\mbox{(stat.)}~^{+1.1}_{-1.2}~\mbox{(syst.)}$~$\mu$b  at $W_0=71.7$~GeV and
$a=0.16\pm 0.06~\mbox{(stat.)}~^{+0.11}_{-0.15}~\mbox{(syst.)}$. The value of 
$a$ is consistent with the value expected for a ``soft" pomeron, 
$a\simeq 0.22$ (see e.g.~\cite{dola}). The systematic uncertainties were
determined by repeating the fit to the cross section obtained after each 
systematic check. The differences between the values of
$\sigma_{\gamma p \to \rho^0 p}(W_0)$ and $a$ thus found and the nominal value 
were added in quadrature. 
The dominant contribution to the systematic uncertainty on
$a$ is that due to the trigger efficiency since its effect in different
$W$ bins is not correlated; conversely, the effects of the uncertainty of 
the proton-dissociative
background contamination and that of the procedure to extract 
the resonant part of the cross section are the same in all $W$ bins.

\subsubsection{Differential cross section $d\sigma/d|t|$}
\label{t_el}

\begin{figure}
\vspace{-4cm}
\begin{center}
\leavevmode
\hbox{%
\epsfxsize = 17cm
\epsffile{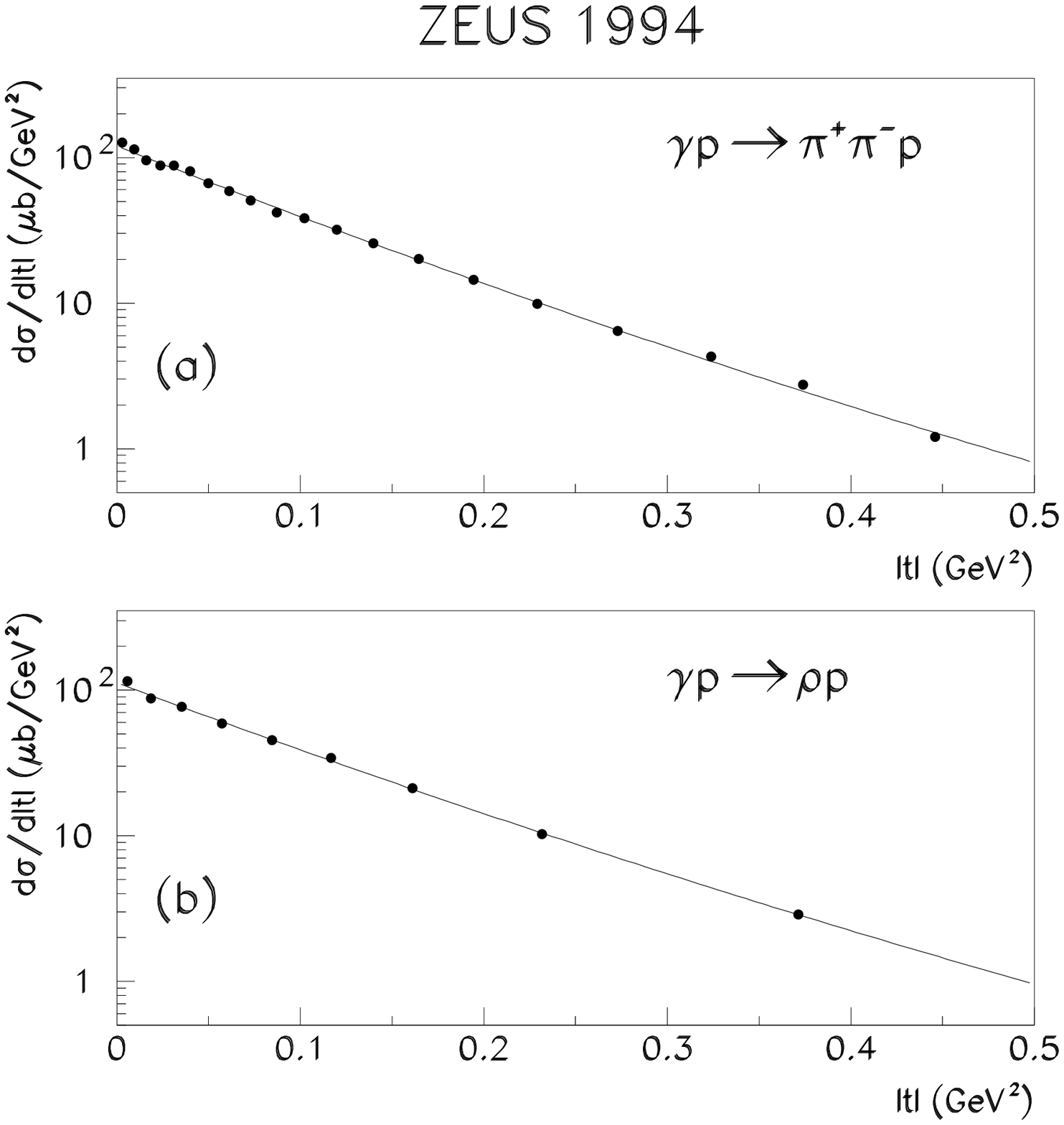}}
\end{center}
\vspace{1cm}
\caption{
(a) The differential cross section $d\sigma/d|t|$ for the process 
$\gamma p \rightarrow \pi^+ \pi^- p$ in the kinematic range 
$0.55<M_{\pi\pi}<1.2$~GeV and $50 <W<100$~GeV.
(b) The differential cross section $d\sigma/d|t|$ for the process $\gamma p 
\rightarrow \rho^0 p$ in the kinematic range 
$2M_{\pi}<M_{\pi\pi}<M_{\rho}+5\Gamma_0$  and $50 <W<100$~GeV.
The continuous lines in (a) and (b) represent the results of the fits with the 
functional forms~(\protect\ref{eqexpt}) and (\protect\ref{eqext}),
respectively. 
The error bars represent only the statistical uncertainties and are smaller 
than the size of the symbols.
}
\label{fig_pt2_t}
\end{figure}

Figure~\ref{fig_pt2_t}a shows the differential cross section 
${d \sigma}/{d|t|}$ for the process 
$\gamma p \rightarrow \pi^+\pi^- p$ in the kinematic range 
$0.55<M_{\pi \pi}<1.2$~GeV and $50<W<100$~GeV. 
The cross section exhibits the exponential fall characteristic of 
diffractive processes. A fit to the form
\begin{eqnarray}
\frac{ d\sigma}{d|t|} = A_{\pi\pi} e^{-b_{\pi\pi} |t| + c_{\pi\pi} t^2}
\label{eqexpt}
\end{eqnarray}

\noindent 
was performed. The fitted values of $b_{\pi\pi}$ and $c_{\pi\pi}$ are
$11.4 \pm 0.3$~(stat.)~$^{+0.3}_{-0.5}$~(syst.)~GeV$^{-2}$ and 
$2.8  \pm 0.7$~(stat.)~$^{+1.2}_{-1.8}$~(syst.)~GeV$^{-4}$,
respectively. 
The main contribution to the systematic errors
is the uncertainty of the acceptance.

\begin{figure}
\vspace{-2.0cm}
\begin{center}
\leavevmode
\hbox{%
\epsfxsize = 15cm
\epsffile{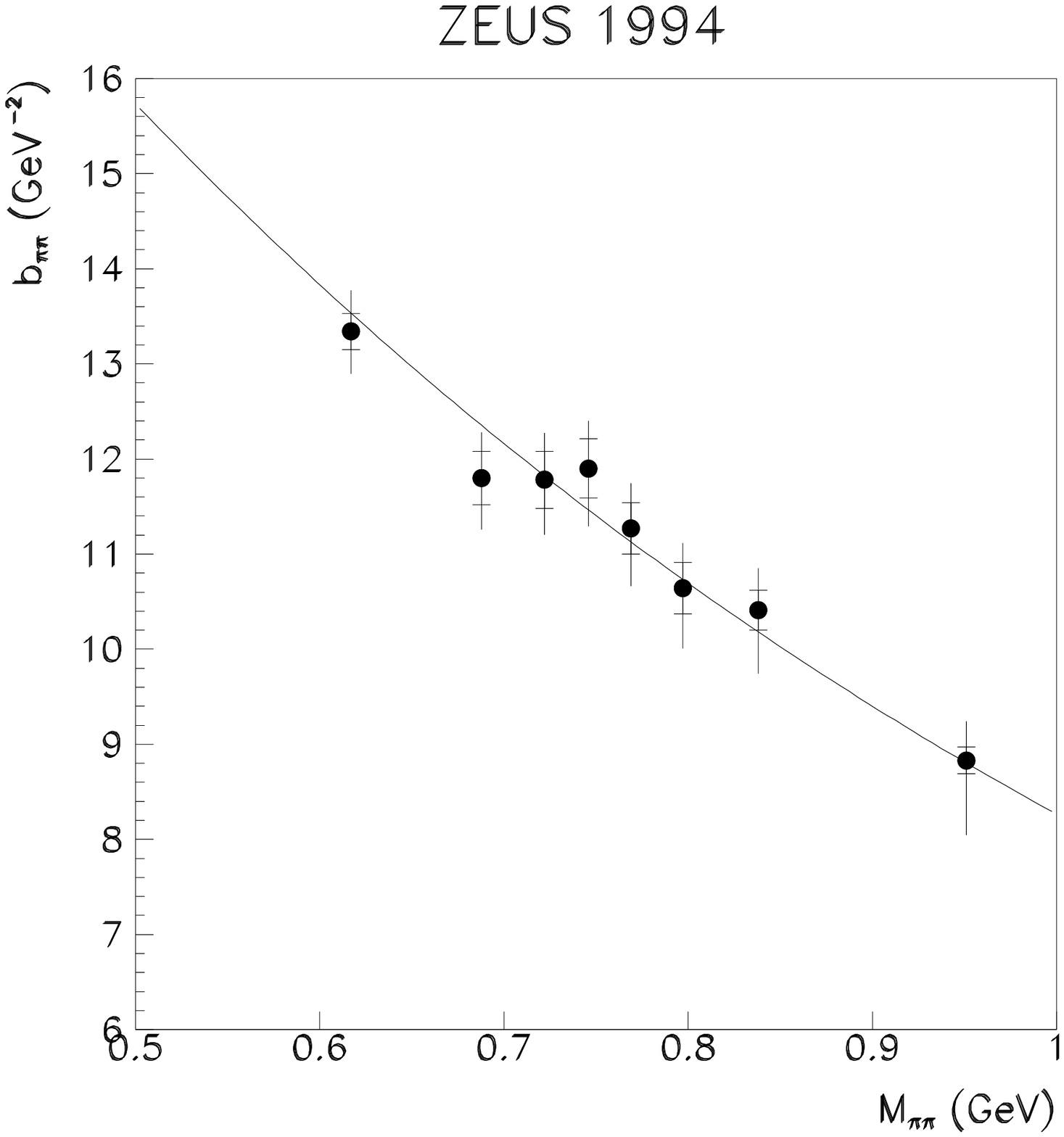}}
\end{center}
\vspace{1.0cm}
\caption{The slope $b_{\pi\pi}$ resulting from a fit of 
equation~(\protect\ref{eqexpt}) to the $t$ distribution for the reaction
$\gamma p \rightarrow \pi^+ \pi^- p$ in different mass bins. 
The kinematic range is 
$50<W<100$~GeV and $|t|<0.5$~GeV$^2$. 
The continuous line indicates the effective expectation of the
S\"oding model as implemented in the 
calculation by Ryskin and Shabelski~\protect\cite{misha_yuly} 
(cf. section~\protect\ref{sec_pi_p}). 
The inner bars 
indicate the statistical uncertainty and the outer ones the 
statistical and systematic uncertainties summed in quadrature.
}
\label{fig_b} 
\end{figure}

In Fig.~\ref{fig_b} the 
slope $b_{\pi\pi}$ resulting from a fit of equation~(\ref{eqexpt}) 
in different mass bins is shown; in this case $c_{\pi\pi}$ was kept 
fixed at the value 2.8~GeV$^{-4}$.
The rapid decrease of $b_{\pi\pi}$ with increasing mass is consistent 
with the results of earlier measurements (cf. e.g.~\cite{bauer}) 
and effectively is expected in the
S\"oding model~\cite{soeding,misha_yuly} 
as the continuous curve in Fig.~\ref{fig_b} shows. The
way the curve was obtained is discussed in section~\ref{sec_pi_p}.

In order to determine $d\sigma/d|t|$ for the resonant process
$\gamma p \rightarrow \rho^0 p$, the 
mass fits with eq.~(\ref{equsod}) were carried out in each $|t|$ bin
(with $M_{\rho}$ and $\Gamma_0$ fixed to the values of 
table~\ref{tab_mass_fit})
and the resonant part of the cross section extracted as a function of 
$|t|$ and integrated over the range 
$2M_\pi < M_{\pi \pi} < M_\rho + 5 \Gamma_0$.
The cross section $d\sigma/d|t|$ thus obtained 
is plotted in Fig.~\ref{fig_pt2_t}b, where the result of 
the fit with the function 
\begin{eqnarray}
\frac{ d\sigma}{d|t|} = A_{\rho} e^{-b_{\rho} |t| + c_{\rho} t^2}
\label{eqext}
\end{eqnarray}

\noindent 
is also shown. The parameters 
of the fit are $b_{\rho}= 10.9 \pm 0.3$~(stat.)~$^{+1.0}_{-0.5}$~(syst.)~GeV$^{-2}$ and 
$c_{\rho}= 2.7 \pm 0.9$~(stat.)~$^{+1.9}_{-1.7}$~(syst.)~GeV$^{-4}$.
The larger systematic uncertainty of $b_{\rho}$ with respect
to that of $b_{\pi\pi}$ reflects the sensitivity to the procedure used to extract
the resonant part of the cross section.

\begin{figure}
\vspace{-2.0cm}
\begin{center}
\leavevmode
\hbox{%
\epsfxsize = 15cm
\epsffile{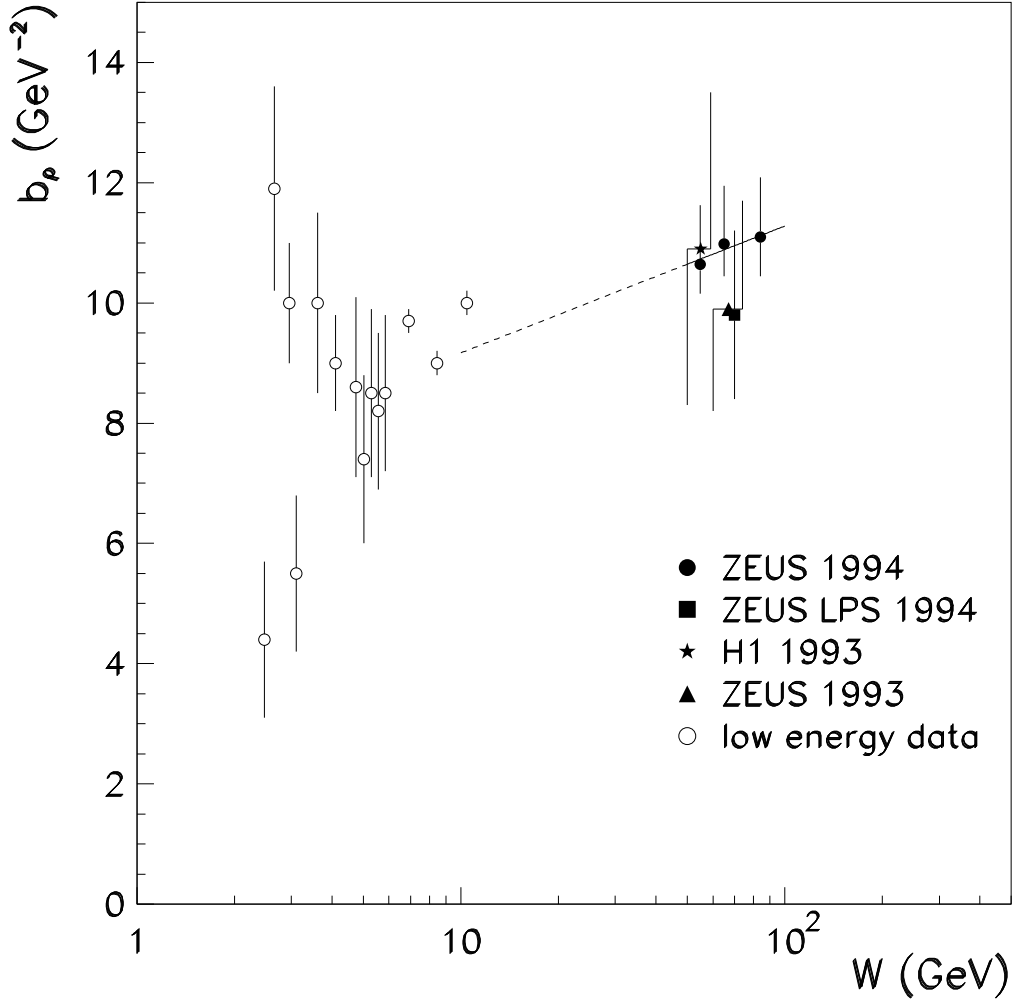}}
\end{center}
\vspace{1.0cm}
\caption{The slope $b_{\rho}$ for the
elastic reaction $\gamma p \rightarrow \rho^0 p$ in the kinematic region
$50<W<100$~GeV and $|t|<0.5$~GeV$^2$ as a function of $W$ together with 
the other recent results from HERA~\protect\cite{rho93,h1rho,lps_rho} and a 
compilation of low energy 
data~\protect\cite{omega,jones,berger,gladding,ballam0}. The continuous 
line shows the result of the fit discussed in the text; the 
extrapolation of the fit to the low $W$ region is indicated by the 
dashed line.
The error bars of the HERA data indicate the statistical and 
systematic uncertainties summed in quadrature. For the points 
at the same value of $W$, the error bars have been offset.
}
\label{b_vs_W} 
\end{figure}

Finally the $|t|$ distribution was studied in three different $W$ bins; the parameter
$c_{\rho}$ was fixed to 2.7~GeV$^{-4}$. 
Table~\ref{tab_b_rho} and Fig.~\ref{b_vs_W} show the values of the slope $b_{\rho}$ as 
a function of $W$ together with the other recent results from 
HERA~\cite{rho93,h1rho,lps_rho} and a partial compilation of low energy 
data~\cite{omega,jones,berger,gladding,ballam0} (cf. Fig.~9 of 
ref.~\cite{rho93}). 

\begin{table}[t]\centering
\begin{tabular}{cc}
\hline \hline
$\langle {\rm W} \rangle$ [GeV] & $b_{\rho}$ [GeV$^{-2}$] \\ 
\hline
 55 & 10.6 $\pm$ $0.2~(\mbox{stat.}) ^{+1.0}_{-0.4}~(\mbox{syst.}) $ \\ 
\hline
 65 & 11.0 $\pm$ $0.2~(\mbox{stat.}) ^{+1.0}_{-0.5}~(\mbox{syst.}) $ \\ 
\hline
 84 & 11.1 $\pm$ $0.1~(\mbox{stat.}) ^{+1.0}_{-0.6}~(\mbox{syst.}) $ \\ 
\hline \hline
\end{tabular}
\caption{$b_{\rho}$ as a function of $W$.}
\label{tab_b_rho}
\end{table}

A fit of the form 
$b_{\rho}(W)=b_{\rho}(W_0)+2 \alphapom^{\prime}\ln{(W/W_0)^2}$, with
$W_0=71.7$~GeV,
to the present data alone yields $\alphapom^{\prime}= 
0.23~\pm 0.15~\mbox{(stat.)}~^{+0.10}_{-0.07}~\mbox{(syst.)}$~GeV$^{-2}$.
The systematic uncertainty was determined by 
repeating the fit to the $b$ values as modified by the effect
of each individual systematic uncertainty; the differences between the values 
of $\alphapom^{\prime}$ thus found and the nominal value were added in quadrature. 
The present result is 
consistent with $\alphapom^{\prime}= 0.25$~GeV$^{-2}$ obtained from fits 
to data on soft hadronic processes~\cite{dola}. Such a dependence of 
$b_{\rho}$ on $W$ is expected to be valid for $W \gsim 5$-10~GeV~\cite{dola}.

\subsubsection{Decay angular distributions}

The angular distributions of the decay pions allow one to determine 
the $\rho^0$ spin 
density matrix elements. They were determined in the helicity frame,
where
the dependence on $\theta_h$ 
and $\varphi_h$ can be written as~\cite{schilling-wolf}:

\begin{eqnarray}
\frac{1}{\sigma}\frac{d\sigma}{d\cos\theta_h d\varphi_h} & = &
\frac{3}{4 \pi} [\frac{1}{2}\left(1-r_{00}^{04}\right)+
\frac{1}{2}\left(3r_{00}^{04}-1\right)\cos^2{\theta_h}
-\sqrt{2}\Re{e[r^{04}_{10}]}\sin{2\theta_h}\cos{\varphi_h}\nonumber \\
& & -r_{1-1}^{04}\sin^2{\theta_h}\cos{2\varphi_h}], 
\label{ang_dist}
\end{eqnarray}

\noindent
with $r_{00}^{04}$, $r_{10}^{04}$ and $r_{1-1}^{04}$ the 
$\rho^0$ spin density matrix elements. The element $r_{00}^{04}$ 
represents the probability that the produced $\rho^0$ meson has
helicity 0; $r_{1-1}^{04}$ is related to
the size of the interference between the helicity non-flip and 
double flip amplitudes, while $\Re{e[r^{04}_{10}]}$ is related to the 
interference between the helicity non-flip and single flip amplitudes.
If $s$-channel helicity conservation (SCHC) holds, 
$r_{1-1}^{04}$ and $\Re{e[r^{04}_{10}]}$ should be zero; $r_{00}^{04}$ 
should also be small because in the kinematic range of the present data
the incoming photons are mostly transverse.

\begin{figure}
\vspace{-2.0cm}
\begin{center}
\leavevmode
\hbox{%
\epsfxsize = 15cm
\epsffile{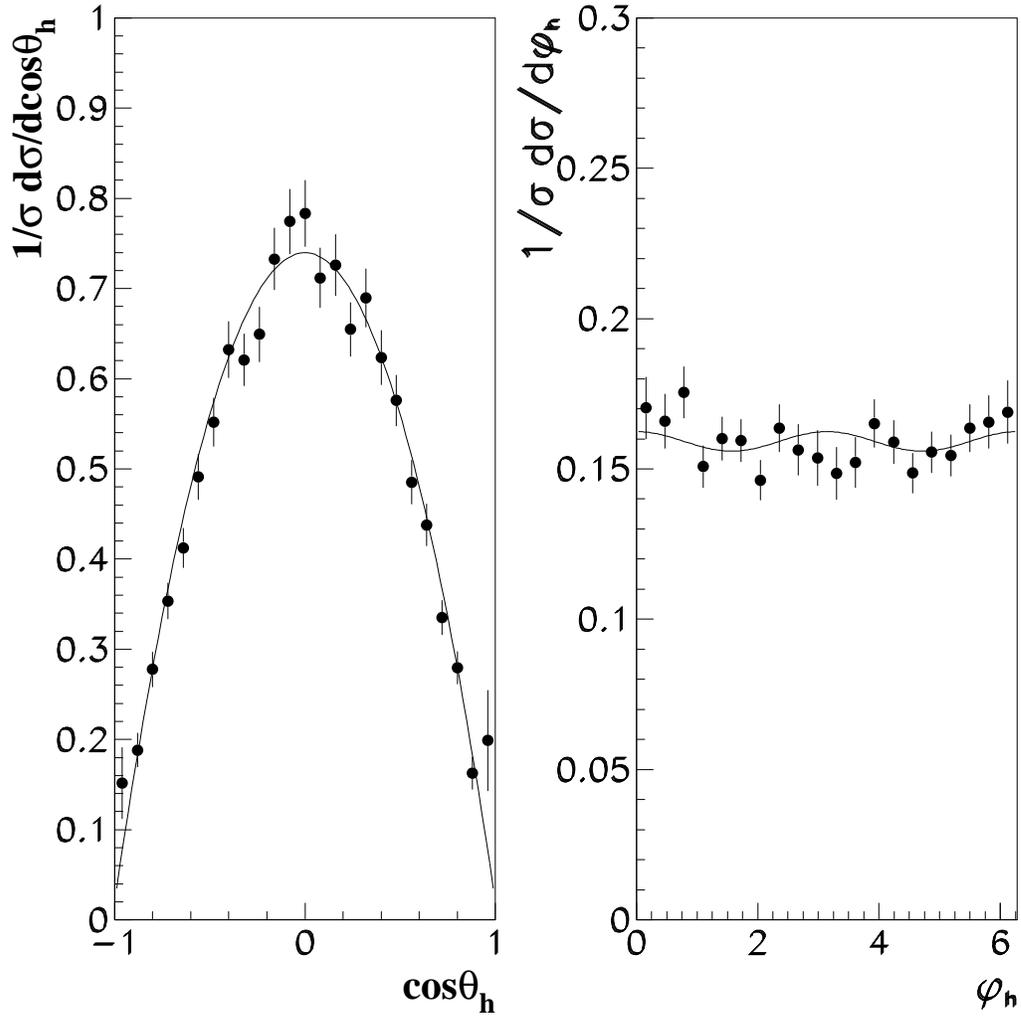}}
\end{center}
\vspace{1.0cm}
\caption{
The differential distributions $(1/\sigma) (d\sigma/d\cos\theta_h)$ and 
$(1/\sigma) (d\sigma/d\varphi_h)$ for the
reaction $\gamma p \rightarrow  \pi^+ \pi^- p $ in the kinematic region
$50<W<100$~GeV and $|t|<0.5$~GeV$^2$. The continuous lines represent the results of the 
fit discussed in the text. Only statistical errors are shown. 
}
\label{angular_dist}
\end{figure}

Figure~\ref{angular_dist} shows the acceptance corrected 
$\theta_h$ and $\varphi_h$ distributions for the process
$\gamma p \rightarrow \pi^+\pi^- p$. As discussed above 
(cf. Fig.~\ref{ratio_pdiss}), their shape is consistent with 
being the same for elastic and proton-dissociative events.

A two-dimensional least-squares fit 
of equation~(\ref{ang_dist}) to the acceptance 
corrected $\cos\theta_h$ and $\varphi_h$ distributions yields 
$r_{0 0}^{0 4} = 0.01 \pm 0.01~\mbox{(stat.)} \pm 0.02~\mbox{(syst.)}$, 
$r_{1 -1}^{0 4}=-0.01 \pm 0.01~\mbox{(stat.)} \pm 0.01~\mbox{(syst.)}$
and 
$\Re{e[r^{04}_{10}]}=0.01 \pm 0.01~\mbox{(stat.)} \pm 0.01~\mbox{(syst.)}$. 
The result of the fit is shown in Fig.~\ref{angular_dist}.
The $\chi^2/ndf$ of the fit is 225/215.
A moment analysis gives similar values. The systematic uncertainties are
dominated by the error of the acceptance. The 
two-dimensional $\theta_h$, $\varphi_h$ distribution was 
not corrected for the non-resonant and interference contributions, 
which however appear to have the same $\cos{\theta_h}$ and $\varphi_h$
dependence as the resonant process (cf. Fig.~\ref{fig_basod_1}).

The present results indicate that in the kinematic range studied
the $\rho^0$ mesons are produced predominantly with helicity $\pm 1$. 
In addition our data are consistent with $s$-channel helicity 
conservation. 

\begin{figure}
\vspace{-2.0cm}
\begin{center}
\leavevmode
\hbox{%
\epsfxsize = 15cm
\epsffile{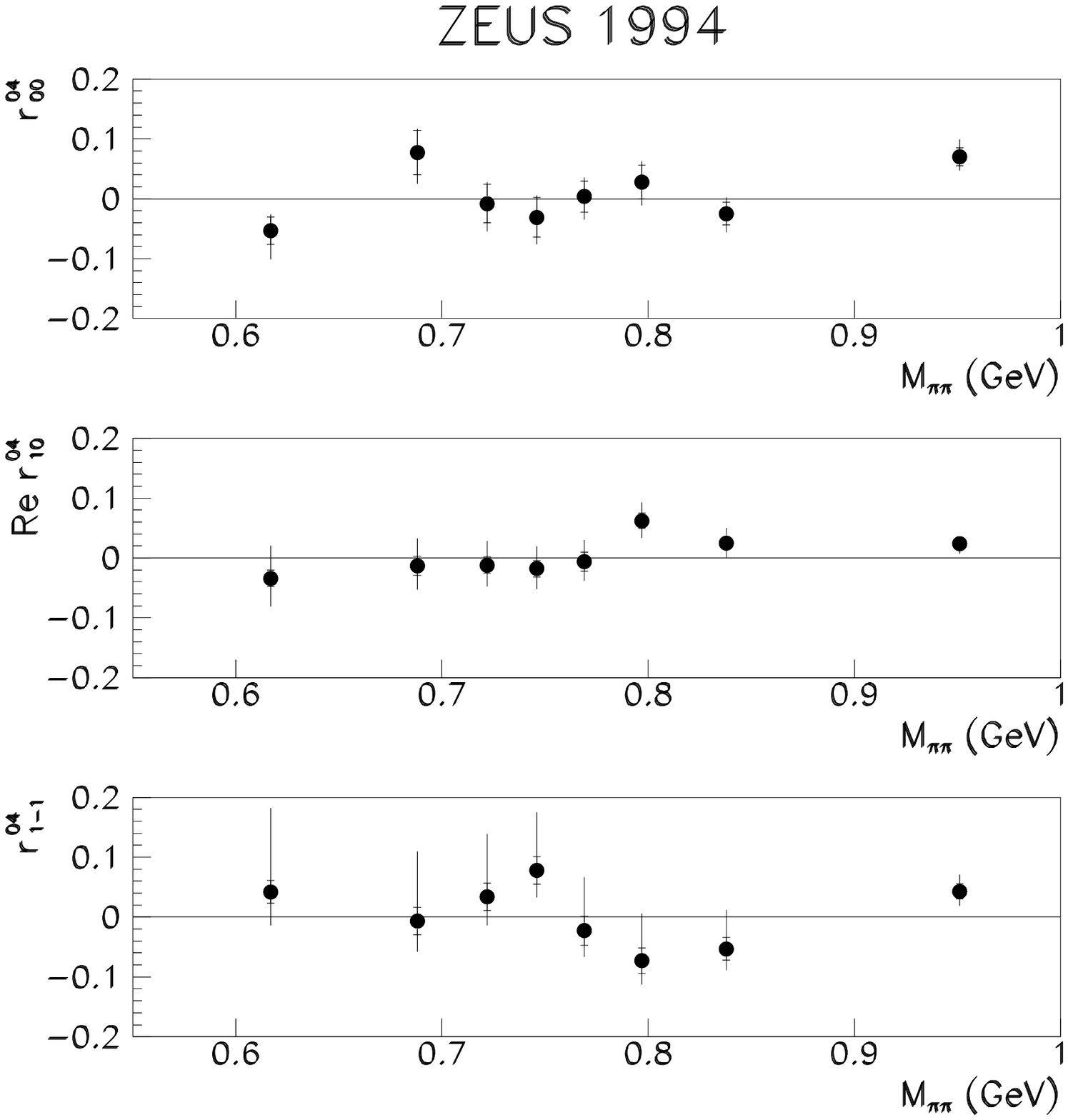}}
\end{center}
\vspace{1.0cm}
\caption{The results for $r_{00}^{04}$, $\Re{e[r^{04}_{10}]}$ and 
$r_{1-1}^{04}$ as a function of $M_{\pi \pi}$ for the reaction
$\gamma p \rightarrow \pi^+ \pi^- p $ in the kinematic range 
$0.55<M_{\pi\pi}<1.2$~GeV, $50<W<100$~GeV and $|t|<0.5$~GeV$^2$. 
The inner bars indicate the statistical uncertainty and the outer ones the 
statistical and systematic uncertainties summed in quadrature.
}
\label{r_vs_mpipi}
\end{figure}

\begin{figure}
\vspace{-2.0cm}
\begin{center}
\leavevmode
\hbox{%
\epsfxsize = 15cm
\epsffile{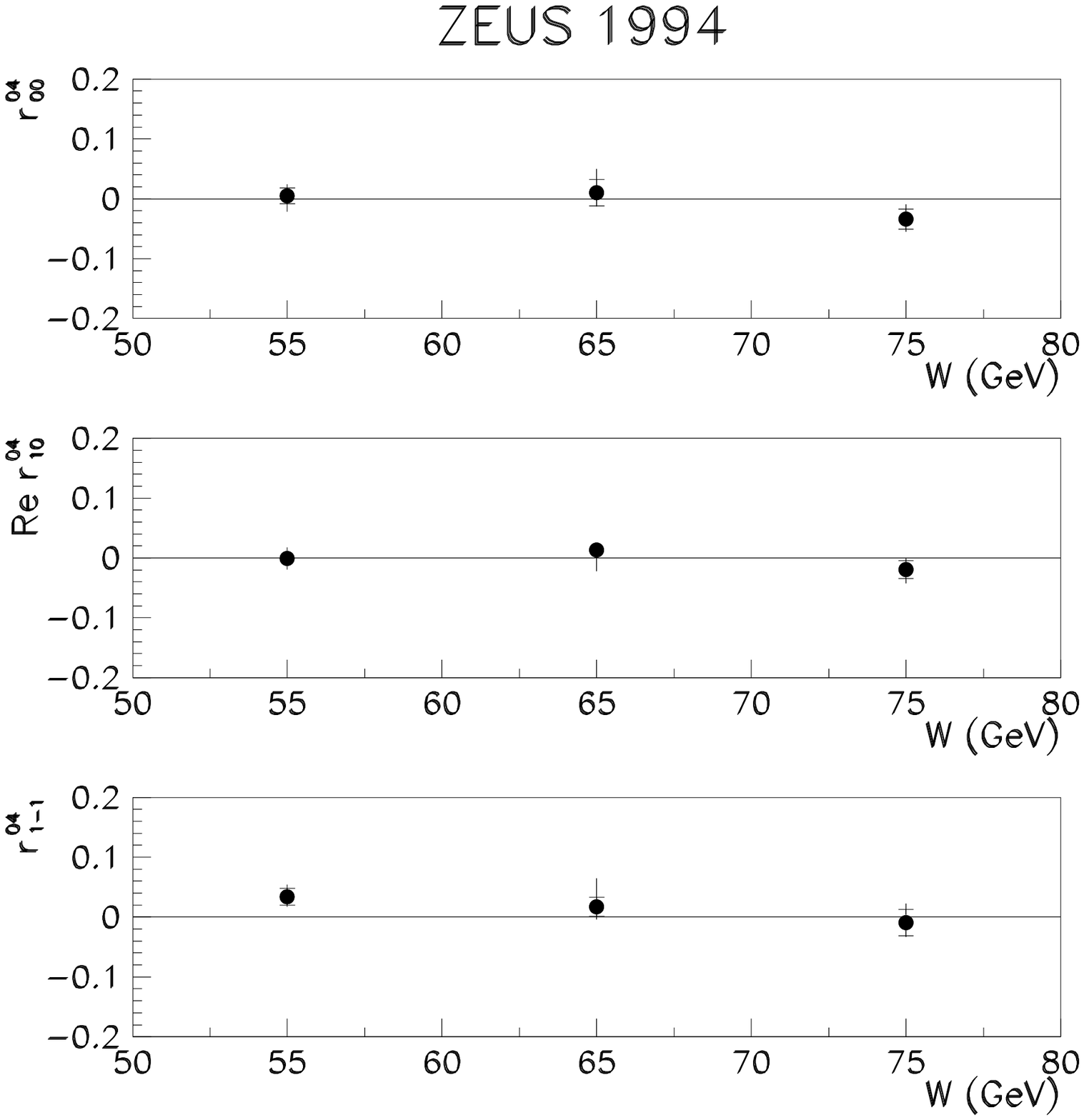}}
\end{center}
\vspace{1.0cm}
\caption{The results for $r_{00}^{04}$, $\Re{e[r^{04}_{10}]}$ and 
$r_{1-1}^{04}$ as a function of $W$ for the reaction
$\gamma p \rightarrow \pi^+ \pi^- p $ in the kinematic range 
$0.55<M_{\pi\pi}<1.2$~GeV, $50<W<80$~GeV and $|t|<0.5$~GeV$^2$. 
The inner bars indicate the statistical uncertainty and the outer ones the 
statistical and systematic uncertainties summed in quadrature.
}
\label{r_vs_W}
\end{figure}

The two-dimensional fit described above was repeated in different 
$M_{\pi \pi}$ and $W$ bins. The results found for $r_{00}^{04}$, 
$\Re{e[r^{04}_{10}]}$ and $r_{1-1}^{04}$ are plotted as a function of 
$M_{\pi \pi}$ in Fig.~\ref{r_vs_mpipi}; the data do not indicate any
dependence on $M_{\pi \pi}$. It should be noted that 
in some 
models (see e.g.~\cite{misha_yuly_2}), for finite values 
of $Q^2$ ($Q^2 \gsim 1$~GeV$^2$), $r^{04}_{00}$ is expected to 
vary with $M_{\pi\pi}$. A variation at large values of $M_{\pi\pi}$,
$M_{\pi\pi}\gsim 0.9$, was observed in fixed target 
photoproduction experiments~\cite{ballam}.
The results are also independent of $W$ as shown in Fig.~\ref{r_vs_W}; 
for this study the $W$ range was restricted to $W<80$~GeV since at large 
$W$ the two-dimensional acceptance in the $\cos{\theta_h,\varphi_h}$ plane 
is rapidly varying. Here again, the comparison of 
these results with those obtained for $r_{1-1}^{04}$ and $\Re{e[r^{04}_{10}]}$ 
by the low energy experiments (cf. e.g.~\cite{ballam0}) confirms the
lack of $W$ dependence for these elements.
Further investigations, not presented here, show that 
if SCHC and natural parity exchange in the $t$-channel are assumed, then
$r_{00}^{04}$, $\Re{e[r^{04}_{10}]}$ and $r_{1-1}^{04}$ appear
independent also of $t$, in the $t$ range studied here~\cite{thesis_dirk}.

\section{Proton-dissociative $\rho^0$ photoproduction}
\label{p_diss_results}

\subsection{$M_{\pi\pi}$, $W$, $\cos{\theta_h}$, $\varphi_h$ distributions}
\label{p_diss_massetc}

As discussed earlier (section~\ref{background_elastic}),
proton-dissociative events selected with the 
PRT1 or the FCAL as described in section~\ref{selection_inelastic} 
have the same 
$M_{\pi\pi}$, $W$, $\cos{\theta_h}$ and $\varphi_h$ dependence
as the elastic events selected by the cuts of 
section~\ref{selection_elastic} (which contain only a contamination 
$R_{diss}=20\%$ from proton-dissociative events). This was deduced from the
fact that the ratios
of the uncorrected $M_{\pi\pi}$, $W$, $\cos{\theta_h}$ and $\varphi_h$ 
distributions for the proton-dissociative sample and the sample 
obtained with the elastic cuts are consistent with being flat, as
Fig.~\ref{ratio_pdiss} shows.

This result supports the hypothesis of factorisation of the diffractive 
vertices. As discussed in~\cite{reviews}, given the dissociative reaction
$ha \rightarrow Na$ and the elastic one $ha \rightarrow ha$, with $h$ 
and $a$ hadrons, factorisation implies 

\begin{eqnarray}
\frac{d^2\sigma_{diss}/dtd(M_N^2/s_{ha})} {d^2\sigma_{el}/dt}=f(s,M_N^2/s_{ha},t),
\label{factorisation}
\end{eqnarray}

\noindent
i.e. at given $s_{ha}$, $M_N^2$ and $t$, the ratio of the diffractive 
dissociation cross section to the elastic cross section is a constant 
independent of hadron $a$; here $s_{ha}$ indicates the square of the centre-of-mass
energy of the $ha$ system.

The results presented in sections~\ref{t_inel} and~\ref{elastic_vs_inelastic}
were obtained for the production of pion pairs in the range 
$0.55<M_{\pi\pi}<1.2$~GeV and not for the resonant process. This was done
because of the limited statistics. 

\subsection{$|t|$ distribution}
\label{t_inel}

\begin{figure}
\vspace{-2.0cm}
\begin{center}
\leavevmode
\hbox{%
\epsfxsize = 15cm
\epsffile{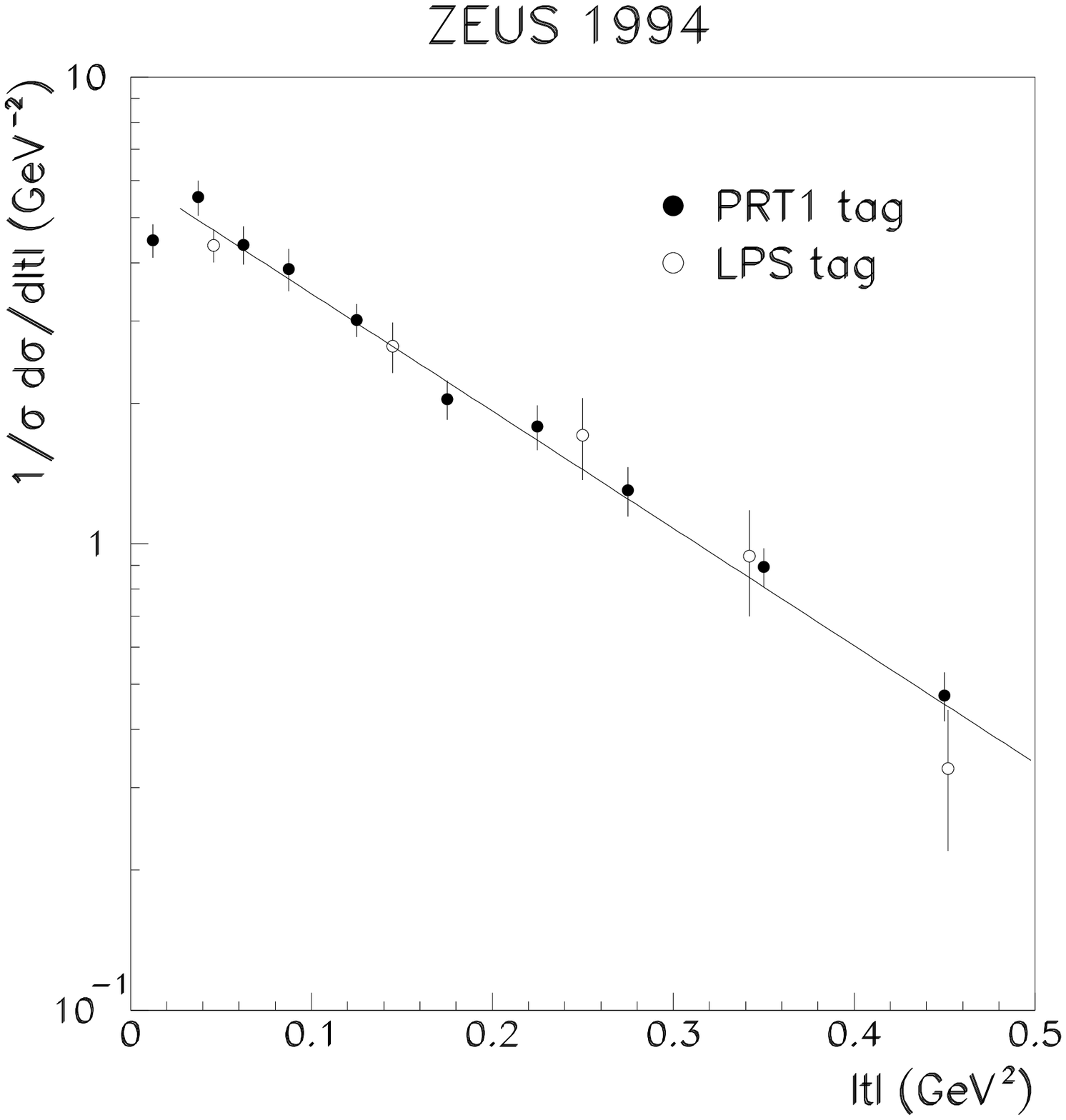}}
\end{center}
\vspace{1.0cm}
\caption{
$t$ distribution for the reaction $\gamma p \rightarrow \pi^+ \pi^- N$ tagged 
with the PRT1 (full symbols) and with the LPS ($x_L<0.98$, open symbols) 
in the region $0.55<M_{\pi\pi}<1.2$~GeV, $50<W<100$~GeV, 
$|t|<0.5$~GeV$^2$ and $(M_p+M_{\pi})^2<M_N^2<0.1 W^2$. 
The dip at low $|t|$ is a consequence of $t_{\min}$ being non-zero at 
large values of $M_N$. Only statistical errors are shown.
The line shows the result of the fit to the PRT1 points described in 
the text. }
\label{fig_in_t}
\end{figure}

As discussed in section~\ref{background_elastic},
the $p_{T\rho}^2$ and hence the $|t|$ distribution for proton-dissociative events 
is shallower than for elastic events.
The acceptance corrected $|t|$ distribution for the reaction
$\gamma p \rightarrow \pi^+ \pi^- N$ obtained with the PRT1 
tagged events is shown in Fig.~\ref{fig_in_t} (solid symbols). 
The continuous line represents the result of a fit with an exponential 
function of the form $A e^{-b_{diss}|t|}$ in the range $0.025<|t|<0.5$~GeV$^2$
and corresponds to a $t$-slope 
$b_{diss} = 5.8 \pm 0.3~(\mbox{stat.}) 
\pm 0.5~(\mbox{syst.})$~${\rm GeV}^{-2}$ for 
the kinematic range $50 < W < 100$ GeV and 
$(M_p+M_{\pi})^2<M_N^2<0.1 W^2$; the upper limit of $M_N$ 
($M_{N_{\max}}=\sqrt{0.1 W^2_{\max}}\approx 30$~GeV) 
was chosen 
following refs.~\cite{chapin,reviews} and corresponds to the region where
diffractive interactions dominate. A fit with a function of the form 
$A\exp{(-b^{\prime}_{diss}|t|+c^{\prime}_{diss}t^2)}$ gives
$b^{\prime}_{diss} = 6.6\pm 1.1~(\mbox{stat.})$~${\rm GeV}^{-2}$  
and 
$c^{\prime}_{diss} = 1.8 \pm 2.4~(\mbox{stat.})$~${\rm GeV}^{-4}$.
If the analysis is repeated for
$(M_p+M_{\pi})^2<M_N^2<100$~GeV$^2$, which is the region covered by the data,
the $t$-slope for a single exponential is $6.4 \pm 0.3~(\mbox{stat.}) 
\pm 0.6~(\mbox{syst.})$~${\rm GeV}^{-2}$. The dip at low $|t|$ is a 
consequence of $t_{\min}$ being non-zero at large values of $M_N$ 
($|t_{\min}| \approx 0.006$~GeV$^{-2}$ for $M_N=0.1W^2$); it disappears for
$M_N^2<100$~GeV$^2$.

In Fig.~\ref{fig_in_t} the open circles show the distribution for the LPS tagged events in 
the kinematic range $50 < W < 100$ GeV and 
$(M_p+M_{\pi})^2<M_N^2<0.1 W^2$.
A fit of an exponential function to these points yields a slope of
$5.8 \pm 0.5~(\mbox{stat.}) \pm 0.9~(\mbox{syst.})$ ${\rm GeV}^{-2}$,
in agreement with the result found with the PRT1 tagged events.

As mentioned earlier, the $t$ distribution determined both for the
PRT1 and the LPS tagged events is for 
$\gamma p \rightarrow \pi^+ \pi^- N$, not for the resonant process 
$\gamma p \rightarrow \rho^0 N$. 
From the elastic data however one finds that the difference of the $t$-slopes
for the reaction $\gamma p \rightarrow \pi^+ \pi^- p$ and for
$\gamma p \rightarrow \rho^0 p$ is $\simeq 0.5$~GeV$^{-2}$. 

For the result obtained with the PRT1 tagged events, the systematic 
error includes the difference of the result obtained 
with the PRT1 and the FCAL tagged events, as well as the 
sensitivity to the Monte Carlo model used (EPSOFT vs PYTHIA) and 
to the shape of the generated $d\sigma/dM_N^2 \propto (1/M_N)^n$ 
spectrum ($n$ was varied in the range $2.0 < n < 2.4$).
For the result obtained with the LPS tagged
events, the sensitivity to the selection cuts and the fitted $|t|$ range
was also included.

The $t$-slope in proton-dissociative $\rho^0$ photoproduction is thus 
smaller than that for the elastic process by about 5~GeV$^{-2}$. This 
is consistent with the results found for virtual
photons~\cite{h1_pdiss} and with those obtained for hadron-hadron 
collisions~\cite{reviews}-\cite{CDF}; it is also in agreement with
theoretical estimates (cf. e.g.~\cite{kolya}).

\subsection{Ratio of the elastic to the proton-dissociative $\rho^0$ photoproduction cross sections}
\label{elastic_vs_inelastic}

Because of the large discrepancy in the normalisation of the acceptance
obtained with PYTHIA and EPSOFT for proton-dissociative events, 
the cross section for the process $\gamma p \rightarrow \rho^0 N$ 
was not determined directly from the PRT1, FCAL or low $x_L$ LPS tagged events. 
We instead determined the ratio of the elastic to the proton-dissociative 
$\rho^0$ photoproduction cross sections, $\sigma_{\gamma p \rightarrow 
\rho^0 p}/\sigma_{\gamma p \rightarrow \rho^0 N}$, using the ratio
$R_{diss}$ found by means of the LPS tagged events with 
$x_L>0.98$ (cf. section~\ref{background_elastic}). 

The elastic yield was calculated as $N(1-R_{diss})/\varepsilon_{el}$,
where $N$ is the number of events passing the selection criteria 
presented in section~\ref{selection_elastic} and $R_{diss}$ is the fraction
of proton-dissociative events in this sample (see section~\ref{background_elastic}).
The acceptance $\varepsilon_{el}$ is the one determined with EPSOFT 
for elastic events. The proton-dissociative yield 
was determined from $NR_{diss}/\varepsilon_{diss}^{el-cuts}$,
where $\varepsilon_{diss}^{el-cuts}$ is the acceptance 
for proton-dissociative events
when the criteria used to select the elastic events are
applied (section~\ref{selection_elastic}).
Note that for $\varepsilon_{diss}^{el-cuts}$ EPSOFT and PYTHIA agree in shape and 
normalisation. 
For the kinematic range $50<W<100$~GeV, $|t|<0.5$~GeV$^2$ and
$(M_p+M_{\pi})^2<M_N^2<0.1 W^2$ one obtains 

\begin{eqnarray}
R_{el/p-diss}=\frac{\sigma_{\gamma p \rightarrow \rho^0 p}} 
{\sigma_{\gamma p \rightarrow \rho^0 N}} = \frac{1-R_{diss}}{R_{diss}} 
\frac{\varepsilon_{diss}^{el-cuts}}{\varepsilon_{el}}= 
2.0 \pm 0.2~(\mbox{stat.})~\pm 0.7~(\mbox{syst.}).
\label{el_over_pdiss}
\end{eqnarray}

\noindent
The quoted error is given by the uncertainties on $R_{diss}$ and 
on the acceptance. The result 
was obtained assuming a mass dependence of the type 
$d^2\sigma/dM_N^2 \propto 1/M_N^n$ with $n=2.24$ as measured by 
CDF~\cite{CDF}. Varying $n$ by $\pm 0.2$ induces a change of $\pm 0.3$ 
in $\sigma_{\gamma p \rightarrow \rho^0 p}/\sigma_{\gamma p \rightarrow 
\rho^0 N}$; this is not included in the quoted systematic uncertainty.

The present result is consistent with that found for 
$pp$ collisions at ISR~\cite{isr_results}, $R_{el/p-diss} = 2.08 \pm 0.13$ at a 
centre-of-mass energy of 53 GeV for $M_N^2<0.05 s_{pp}$ ($\sqrt{s_{pp}}$ is the 
proton-proton centre-of-mass energy) and $1.69 \pm 0.11$ for $M_N^2<0.1 s_{pp}$. 
It is also consistent with the result found by the H1 
Collaboration~\cite{h1_pdiss}:
$\sigma_{\gamma p \rightarrow \rho^0 p}/
\sigma_{\gamma p \rightarrow \rho^0 N}= 
1.54~\pm 0.26$~(stat.)~$\pm 0.31$~(syst.) for $7<Q^2<36$~GeV$^2$, 
$60<W<180$~GeV and $M_N^2<0.05W^2$.
The ZEUS and H1 data together indicate 
that the ratio $\sigma_{\gamma p \rightarrow \rho^0 p}/
\sigma_{\gamma p \rightarrow \rho^0 N}$ is not a strong function of the photon 
virtuality. Our result in conjunction with the $pp$ data 
and the $ep$ results at
non-zero $Q^2$ supports the hypothesis of factorisation.

\section{A model dependent derivation of the pion-proton cross section}
\label{sec_pi_p}

As discussed earlier, the measured cross section for the process
$\gamma p \rightarrow \pi^+ \pi^- p$ includes the contributions of 
resonant $\rho^0 \rightarrow \pi^+ \pi^-$ production, 
non-resonant $\pi^+ \pi^-$ production and their interference.
Non-resonant $\pi^+ \pi^-$ production can be described by the photon 
fluctuating into a virtual $\pi^+ \pi^-$ pair with one or both pions 
scattering 
elastically off the proton. The amplitude for this process can thus be written in 
terms of the $\pi p$ total cross section $\sigma_{\pi p}$. We 
have extracted this cross 
section in the framework of a recent calculation~\cite{misha_yuly}, based 
on S\"oding's approach~\cite{soeding}, in which $\sigma_{\pi p}$ 
is one of the parameters.

The total $\pi p$ cross section 
was determined by fitting the calculation of ref.~\cite{misha_yuly} 
to the $M_{\pi\pi}$ distribution of Fig.~\ref{fig_mass_tot}. The fit
gives $\sigma_{\pi p}= 31 \pm 2~\mbox{(stat.)}~\pm 3~\mbox{(syst.)}$~mb at 
an average pion-proton centre-of-mass energy 
$\sqrt{s_{\pi p}}\simeq \sqrt{\langle W^2 \rangle/2} \simeq 50$~GeV. The systematic
error reflects the systematic uncertainty of the data. An additional 
uncertainty of approximately 15\% was evaluated 
by repeating the fit with different values of the
parameters of the model. The value of $\chi^2/ndf$ is 23.4/23.

Our result is consistent with the extrapolation of the fits by
Donnachie and Landshoff~\cite{dola} which give $\sigma_{\pi p}=26.6$~mb at 
$\sqrt{s_{\pi p}} = 50$~GeV. 

The predictions of the calculation~\cite{misha_yuly} using the 
fitted value of $\sigma_{\pi p}$ are shown in 
Figs.~\ref{fig_mass_bin},~\ref{fig_basod} and~\ref{fig_b}.
Both the decrease of the skewing with increasing $|t|$ and the 
variation of the $t$-slope with $M_{\pi\pi}$ are well described. To 
obtain the curves shown in Figs.~\ref{fig_basod} and~\ref{fig_b}, 
events were generated with a Monte Carlo program 
based on~\cite{misha_yuly} and were binned as a function of $M_{\pi\pi}$ 
and $t$. The fits performed to the $M_{\pi\pi}$ spectra in the data 
for different $t$ bins (using eq.~(\ref{equross})) and to the $t$ 
distributions for different $M_{\pi\pi}$ bins were repeated for the generated 
events.

\section{Summary and conclusions}

We have presented a high statistics study of $\rho^0$
photoproduction for $50<W<100$~GeV and $|t|<0.5$~GeV$^2$. With respect to 
previous analyses at HERA, the present one features larger statistics and 
reduced systematic uncertainties. The main novel results can be summarised as 
follows:

\begin{itemize}

\item The $\pi^+ \pi^-$ invariant mass spectrum is skewed and the amount of
skewing decreases with increasing $|t|$, consistent with the results from
fixed target experiments.

\item The cross section for resonant $\rho^0$ production, 
$\gamma p \rightarrow \rho^0 p$, 
is $11.2\pm 0.1~\mbox{(stat.)}~^{+1.1}_{-1.2}~\mbox{(syst.)}$~$\mu$b at 
$\langle W\rangle = 71.7$~GeV.
It increases slowly with $W$, exhibiting a 
power-like behaviour of the type $W^a$ with 
$a=0.16\pm 0.06~\mbox{(stat.)}~^{+0.11}_{-0.15}~\mbox{(syst.)}$, 
consistent with $a \simeq 0.22$, the value expected for a ``soft" pomeron.

\item The $t$ distribution for the reaction $\gamma p \rightarrow \pi^+\pi^- p$
is well described by an exponential 
of the form $A_{\pi\pi}\exp{(-b_{\pi\pi}|t|+c_{\pi\pi}t^2)}$. The slope
$b_{\pi\pi}$ decreases rapidly with increasing $M_{\pi\pi}$, again consistent
with the results from fixed target experiments.

The $t$ dependence of the cross section of the reaction
$\gamma p \rightarrow \rho^0 p$ can also be
described by a function of the type  
$A_{\rho} \exp{(-b_{\rho} |t| + c_{\rho} t^2)}$, with 
$b_{\rho}= 10.9 \pm 0.3$~(stat.)~$^{+1.0}_{-0.5}$~(syst.)~GeV$^{-2}$ and 
$c_{\rho}= 2.7 \pm 0.9$~(stat.)~$^{+1.9}_{-1.7}$~(syst.)~GeV$^{-4}$.

A fit with the function 
$b_{\rho}(W)=b_{\rho}(W_0)+2 \alphapom^{\prime}\ln{(W/W_0)^2}$
yields $\alphapom^{\prime}= 
0.23~\pm 0.15~\mbox{(stat.)}$ $^{+0.10}_{-0.07} \mbox{(syst.)}$ GeV$^{-2}$,
consistent with results from elastic hadron-hadron scattering. 

\item The $\rho^0$ spin density matrix elements $r_{00}^{04}$, 
$\Re{e[r^{04}_{10}]}$ and $r_{1-1}^{04}$ were obtained from the 
angular distributions of the 
decay pions in the helicity frame; their values are 
consistent with $s$-channel helicity conservation. No dependence on
$M_{\pi\pi}$ or $W$ is observed.

\item Proton-dissociative $\rho^0$ photoproduction, in which the proton
diffractively dissociates into a system with mass $M_N \lsim 10$~GeV, exhibits
dependences on $M_{\pi\pi}$, $W$, $\cos{\theta_h}$ and $\varphi_h$ consistent
within errors with those of the elastic process. The slope of the $t$ distribution is 
smaller than in the elastic reaction and for $0.55<M_{\pi\pi}<1.2$~GeV
and $(M_p+M_{\pi})^2<M_N^2<0.1 W^2$ is measured to be 
$b = 5.8 \pm 0.3~(\mbox{stat.}) \pm 0.5~(\mbox{syst.})$~${\rm GeV}^{-2}$, 
using the PRT1 tagged events.
In this kinematic region, the ratio of the elastic to proton-dissociative
cross sections is $2.0 \pm 0.2~(\mbox{stat.}) \pm 0.7~(\mbox{syst.})$.

\item A model calculation~\cite{misha_yuly} based on the S\"oding 
approach~\cite{soeding} was fitted to the $M_{\pi\pi}$ 
spectrum for the reaction $\gamma p \rightarrow  \pi^+ \pi^- p$.
The fit yielded 
$\sigma_{\pi p}= 31 \pm 2~\mbox{(stat.)}~\pm 3~\mbox{(syst.)}$~mb at an average 
pion-proton centre-of-mass energy of approximately 50~GeV, consistent 
with the predictions of fits to fixed target $\pi p$ data based 
on the ``soft" pomeron. The model dependent uncertainty was estimated to be
approximately 15\%.

\end{itemize}

In $\rho^0$ photoproduction the photon thus appears to behave like a
vector meson.
The $W$ and $t$ dependences of the cross section
are those expected for elastic hadron-hadron scattering and the object
mediating the interaction appears to be the same pomeron that dominates
the hadron-hadron total cross section. The comparison of the elastic and 
the proton-dissociative reactions suggests that the coupling of the pomeron
to the photon is independent of 
that to the proton, as expected on the basis of 
factorisation. Our results indicate that the $\rho^0$ not only carries the 
quantum numbers of the photon, but also its helicity in the $s$-channel 
system is equal to that of 
the photon.
The skew of the mass shape and its $t$ dependence can also
be understood in terms of soft hadron-hadron interactions and 
simple quantum-mechanical interference between resonant and non-resonant
production of pion pairs.

\section*{Acknowledgements}

We thank the DESY Directorate for their strong support and
encouragement. We acknowledge the assistance of the DESY computing and 
networking staff. 
The remarkable achievements of the HERA machine
group were essential for the successful completion of this
work, and are gratefully appreciated; collaboration with the HERA
group was particularly crucial to the successful installation and
operation of the LPS. We would like to thank B.~Hubbard, a former member of 
ZEUS, for his invaluable contribution to the LPS setting up and
to the 1994 data taking.

It is a pleasure to also thank N.N. Nikolaev, M.G.~Ryskin and 
Y.M.~Shabelski for many stimulating discussions.

\newpage

\newpage

\begin{thebibliography}{99}

\bibitem{bauer} For a review, see e.g. 
T.H. Bauer et al., Rev. Mod. Phys. {\bf 50} (1978) 261.

\bibitem{blechscmidt} H. Blechschmidt et al., Nuovo Cimento {\bf 52A} (1967) 1348.
\bibitem{lanzerotti} L.J. Lanzerotti et al., Phys. Rev. {\bf 166} (1968) 1365.
\bibitem{erbe} ABBHHM Collab., R. Erbe et al., Phys. Rev. {\bf 175} (1968) 1669.
\bibitem{jones} W.G. Jones et al., Phys. Rev. Lett. {\bf 21} (1968) 586.
\bibitem{bulos} F. Bulos et al., Phys. Rev. Lett. {\bf 22} (1969) 490.

\bibitem{mcclellan} G. McClellan et al., Phys. Rev. Lett. {\bf 22} (1969) 374. 
\bibitem{bingham} SBT Collab., H.H. Bingham et al., Phys. Rev. Lett. {\bf 24} (1970) 955.
\bibitem{ballam00} SBT Collab., J. Ballam et al.,  Phys. Rev. Lett. {\bf 24} (1970) 960.
\bibitem{alvensleben} DESY-MIT Collab., H. Alvensleben et al., Nucl. Phys. 
{\bf B18} (1970) 333.
\bibitem{anderson} R. Anderson et al., Phys. Rev. {\bf D1} (1970) 27.
\bibitem{davier} M. Davier et al., Phys. Rev. {\bf D1} (1970) 790.
\bibitem{mcclellan1} G. McClellan et al., Phys. Rev. {\bf D4} (1971) 2683. 
\bibitem{berger} C. Berger et al., Phys. Lett. {\bf 39B} (1972) 659.
\bibitem{eisenberg} SWT Collab., Y. Eisenberg et al., Phys. Rev. {\bf D5} (1972) 15.
\bibitem{park} J. Park et al., Nucl. Phys. {\bf B36} (1972) 404.
\bibitem{ballam0} SBT Collab.,  J. Ballam et al., Phys. Rev. {\bf D5} (1972) 545.
\bibitem{ballam} SBT Collab, J. Ballam et al., Phys. Rev. {\bf D7} (1973) 3150.

\bibitem{gladding} G.E. Gladding et al., Phys. Rev. {\bf D8} (1973) 3721.

\bibitem{alexander}  G. Alexander et al., Nucl. Phys. {\bf B69}  (1974) 445.
\bibitem{barish} B. Barish et al., Phys. Rev. {\bf D9} (1974) 566.
\bibitem{alexander1} G. Alexander et al., Nucl. Phys. {\bf B104} (1976) 397.
\bibitem{struczinski} W. Struczinski et al., Nucl. Phys. {\bf B108} (1976) 45.
\bibitem{egloff} R. M. Egloff et al, Phys. Rev. Lett. {\bf 43} (1979) 657.
\bibitem{omega} OMEGA Photon Collab., D. Aston et al., Nucl. Phys. {\bf B209} (1982) 56.

\bibitem{rho93} ZEUS Collab., M. Derrick et al., Z.Phys. {\bf C69} (1995) 39.
\bibitem{h1rho} H1 Collab., S. Aid et al., Nucl. Phys. {\bf B463} (1996) 3.
\bibitem{lps_rho} ZEUS Collab., M. Derrick et al., Z. Phys. {\bf C73} (1997) 253. 
\bibitem{sakurai} J.J. Sakurai, Phys. Rev. Lett. {\bf 22} (1969) 981.
\bibitem{jpsi} H1 Collab., S. Aid et al., Nucl. Phys. {\bf B472} (1996) 3;\\
ZEUS Collab., J.Breitweg et al., Z. Phys.~{\bf C75} (1997) 215.
\bibitem{h1_pdiss} H1 Collab., C. Adloff et al., Z. Phys. {\bf C75} (1997) 607.
\bibitem{reviews} For reviews, see e.g.\\
G. Alberi and G. Goggi, Phys. Rep. {\bf74} (1981) 1;\\
K. Goulianos, Phys. Rep. {\bf101} (1983) 169;\\
N.P. Zotov and V.A. Tsarev, Sov. Phys. Uspekhi {\bf 31} (1988) 119;\\
G. Giacomelli, Int. J. Mod. Phys. A, vol. 5, no. 2 (1990), 223.
\bibitem{sfm} C. Conta et al., Nucl. Phys. {\bf B175} (1980) 97.
\bibitem{isr_results}
CHLM Collab., M.G. Albrow et al., Nucl. Phys. {\bf B108} (1976) 1;\\
CHLM Collab., J.C.M. Armitage et al., Nucl. Phys. {\bf B194} (1982) 365;\\
N. Amos et al., Phys. Lett. {\bf B120} (1983) 460.
\bibitem{pp_results}
UA4 Collab., M. Bozzo et al., Phys. Lett. {\bf 147B} (1984) 385;\\
UA4 Collab., M. Bozzo et al., Phys. Lett. {\bf 147B} (1984) 392;\\
UA5 Collab., R.E. Ansorge et al., Z. Phys. {\bf C33} (1986) 175;\\
UA4 Collab., D. Bernard et al., Phys. Lett. {\bf B186} (1987) 227.
\bibitem{tevatron}
CDF Collab., F. Abe et al., Phys. Rev. {\bf D50} (1994) 5518.
\bibitem{CDF} CDF Collab., F. Abe et al., Phys. Rev. {\bf D50} (1994) 5535.

\bibitem{soeding} P. S\"{o}ding, Phys. Lett. {\bf 19} (1966) 702.
\bibitem{regge} T. Regge, Nuovo Cimento {\bf 14} (1959) 951;\\
T. Regge, Nuovo Cimento {\bf 18} (1960) 947;\\
see also e.g. P.D.B. Collins, ``An Introduction to 
Regge Theory and High Energy Physics"
(Cambridge University Press, Cambridge 1977).

\bibitem{chapin}
T.J. Chapin et al., Phys.~Rev. {\bf D31} (1985) 17.

\bibitem{misha_yuly} M.G.~Ryskin and Y.M.~Shabelski, preprint hep-ph/9701407 
(1997).
\bibitem{detector_a}
ZEUS Collab., M.~Derrick et al., The ZEUS Detector, Status Report 1993, 
DESY (1993).
\bibitem{detector_b}ZEUS Collab., M.~Derrick et al., Phys.~Lett. {\bf B293} (1992) 465.

\bibitem{vxd}
C.~Alvisi et~al., Nucl.~Instr. Meth. {\bf A305} (1991) 30.

\bibitem{ctd} N. Harnew et al., Nucl.~Instr.~Meth.~{\bf A279} (1989) 290;\\
B. Foster et al., Nucl.~Phys.,~Proc.~Suppl.~{\bf B32} (1993) 181;\\
B. Foster et al., Nucl.~Instr.~Meth.~{\bf A338} (1994) 254.


\bibitem{rtd}
B.~Bock et al., Nucl. Instr. Meth. {\bf A344} (1994) 335.

\bibitem{CAL} M. Derrick et al., Nucl.~Instr.~Meth.~{\bf A309} (1991) 77;\\
A. Andresen et al., Nucl.~Instr.~Meth.~{\bf A309} (1991) 101;\\
A. Caldwell et al., Nucl.~Instr.~Meth.~{\bf A321} (1992) 356;\\
A. Bernstein et al., Nucl.~Instr.~Meth.~{\bf A336} (1993) 23.
        
\bibitem{srtd} A. Bamberger et al., DESY report DESY 97-157 (1997).

\bibitem{mx} ZEUS Collab.,  J. Breitweg et al.,
Z. Phys. {\bf C75} (1997) 421.                                  

\bibitem{lumi}
J.~Andruszk\'ow et al., DESY report DESY~92-066 (1992);\\
ZEUS Collab., M. Derrick et al., Z. Phys. {\bf C63} (1994) 391.

\bibitem{thesis_dirk} D. Westphal, Ph.D. thesis, Hamburg 
University (1997), DESY Internal Report F35D-97-11.

\bibitem{thesis_joern} J. Grosse-Knetter, Ph.D. thesis, Hamburg 
University (1997), DESY Internal Report F35D-97-02.

\bibitem{zeus_omega} ZEUS Collab., M.Derrick et al., 
Z. Phys. {\bf C73} (1996) 73.




\bibitem{michal}
M. Kasprzak, Ph.D. thesis, Warsaw University (1996), 
DESY Internal Report F35D-96-16.

\bibitem{herwig}
G. Marchesini et al., Comp. Phys. Comm. {\bf 67} (1992) 465;\\
B.R. Webber, in ``Proceedings of the Workshop `Physics at HERA'~",
DESY, 29-30 October 1991, editors~W. Buchm\"{u}ller and G. Ingelman, p.~1354;\\
L. Stanco, ibidem, p.~1363.

\bibitem{dipsi} M. Arneodo, L. Lamberti and M.G. Ryskin, 
Comput. Phys. Commun. {\bf 100} (1997) 195. 




\bibitem{pythia}
T. Sj\"{o}strand, Comp. Phys. Comm. {\bf 82} (1994) 74.

\bibitem{kurek} K. Kurek, DESY Report DESY 96-209 (1996) and private 
communications.


\bibitem{pdg} R.M. Barnett et al., Phys. Rev. {\bf D54} (1996) 1.

\bibitem{rosssto} M. Ross and L. Stodolsky, Phys. Rev. {\bf 149} (1966) 1172.


\bibitem{dola} A. Donnachie and P.V. Landshoff, Phys. Lett. {\bf B185} (1987) 403;\\
A. Donnachie and P.V. Landshoff, Nucl. Phys. {\bf B311} (1989) 509;\\
P.V. Landshoff, Nucl. Phys. B (Proc.~Suppl.) {\bf 18C} (1990) 211.

\bibitem{Cudell} J.R. Cudell et al., preprint hep-ph 9601336 (1996).



\bibitem{schilling-wolf} K. Schilling et al., Nucl. Phys. {\bf B15} (1970) 
397;\\
K. Schilling and G. Wolf, Nucl. Phys. {\bf B61} (1973) 381.         

\bibitem{misha_yuly_2} M.G.~Ryskin and Y.M.~Shabelski, preprint hep-ph/9704279 
(1997).

\bibitem{kolya} H. Holtmann et al., Z. Phys. {\bf C69} (1996) 297. 

\end{thebibliography}
\end{document}